\documentclass[journal]{IEEEtran}

\usepackage{amsfonts, amssymb, latexsym, amsmath, graphicx, here, cite, color, bbm}
\usepackage[caption=false,font=footnotesize]{subfig}

\newtheorem{thm}{Theorem}
\newtheorem{cor}{Corollary}

\newtheorem{prop}{Proposition}
\newtheorem{defi}{Definition}

\newtheorem{condition}{Condition}
\newtheorem{ex}{Example}

\newcommand\f[1]{\footnotesize{#1}}
\newcommand{\dep}{d_{\epsilon}}
\newcommand{\inn}{\textsf{in}}
\newcommand{\G}{\textsf{G}}

\title{Hypergraph-based Source Codes for Function Computation Under Maximal Distortion} 

\author{ Sourya Basu, Daewon Seo, and Lav R.~Varshney 
	\thanks{ This work was supported in part by the IBM-Illinois Center for Cognitive Computing Systems Research (C3SR), a research collaboration as part of the IBM AI Horizons Network, the Chan Zuckerberg Initiative DAF (2018-182794), an advised fund of Silicon Valley Community Foundation, the Department of Energy (DOE) award (DE-SC0012704), and DGIST Start-up Fund Program of the Ministry of	Science and ICT (2022010019). This work is based in part on papers presented at the 2020 Data Compression Conference (DCC) \cite{BasuSV2020} and the 2020 IEEE International Symposium on Information Theory (ISIT) \cite{BasuSV2020a}.}
	\thanks{ S.~Basu and L.~R.~Varshney are with the Coordinated Science Laboratory and the Department of Electrical and Computer Engineering, University of Illinois Urbana-Champaign, Urbana, IL 61801 USA (e-mail: sourya@illinois.edu; varshney@illinois.edu). D.~Seo is with the Department of Electrical Engineering and Computer Science, DGIST, Daegu 42988, South Korea (e-mail: dwseo@dgist.ac.kr).}
}

\begin{document}

\maketitle

\begin{abstract}
	This work investigates functional source coding problems with maximal distortion, motivated by approximate function computation in many modern applications. The maximal distortion treats imprecise reconstruction of a function value as good as perfect computation if it deviates less than a tolerance level, while treating reconstruction that differs by more than that level as a failure. Using a geometric understanding of the maximal distortion, we propose a hypergraph-based source coding scheme for function computation that is constructive in the sense that it gives an explicit procedure for finding optimal or good auxiliary random variables. Moreover, we find that the hypergraph-based coding scheme achieves the optimal rate-distortion function in the setting of coding for computing with side information and achieves the Berger-Tung sum-rate inner bound in the setting of distributed source coding for computing. It also achieves the El Gamal-Cover inner bound for multiple description coding for computing and is optimal for successive refinement and cascade multiple description problems for computing. Lastly, the benefit of complexity reduction of finding a forward test channel is shown for a class of Markov sources.
\end{abstract}


\section{Introduction} \label{sec:intro}
A primary purpose of many modern communication and storage systems, such as cloud infrastructures, distributed systems, and artificial intelligence systems, is to compute a certain function of data or make a decision from data. Since the output of such a function has less information than the input data, compressing data for function computation can save communication and storage budget more than compressing to maintain fidelity of the data itself. Fundamental limits of such compression for function computation have been studied in information theory over the last few decades under the name of \emph{coding for computing} or \emph{functional compression} \cite{KornerM1979, OrlitskyR2001, DoshiSME2010, MisraGV2011, FeiziM2014}.

Modern computing applications in the era of large-scale data have seen the size and dimensionality of data exceed hardware computing capabilities. In such an environment, computing the exact function from data is possible only with very high (or infeasible) hardware complexity and energy. \emph{Approximate computing} has recently emerged as an alternative approach to simplify hardware complexity, where a certain error tolerance in function computation is allowed \cite{HanO2013, XuMK2016}. For example in image processing, having blurred compressed images is usually acceptable for downstream tasks such as image labeling.

In this work, we formulate functional compression problems under the considerations of approximate computing, where the reconstruction of function values is as useful as perfect reconstruction if recovered values are close to the exact ones within a certain tolerance level. Otherwise, the reconstruction is treated as completely useless. Note that canonical distortion measures do not capture this aspect of approximate computing. For instance, quadratic distortion penalizes less for a smaller deviation but does not consider reconstruction with a small deviation to be perfect. Hamming distortion, commonly used for a discrete subset of a non-metric space, only determines whether or not reconstructed values exactly coincide, thus not applicable for approximate computing that relies on a distance measure. Therefore, to capture the nature of approximate computing, we introduce the \emph{maximal} distortion. For point-to-point functional compression with a function $f:\mathcal{X} \mapsto \mathcal{Z}$, the goal of compression is to send a concise description so that a decoder recovers $f(x)$, instead of $x$. Then, the maximal distortion for $f$ with tolerance level $\epsilon \ge 0$ is formally defined as
\begin{align*}
	\dep(x, z) = \dep(x, z ; f) = \mathbbm{1}_{ \{ \| f(x) - z \| > \epsilon \} },
\end{align*}
where $\| \cdot \|$ is the norm of the image space and $\mathbbm{1}_{\{ \cdot \}}$ is the indicator function. We seek a covering up to functional distortion $\epsilon$, which generalizes Posner and Rodemich's study for $f(x)=x$ \cite{PosnerR1971}.

As one can see, the maximal distortion $\dep(x, z)$ is still a special instance of a general distortion measure in rate-distortion theory, implying that existing non-functional rate-distortion results immediately apply. However, many existing results in source coding are written using auxiliary random variables and provide little guidance on how to find them. This difficulty is more severe in coding for computing because the target function $f$ often can be approximated by any $z \in \mathcal{Z}$ (e.g., $\mathcal{Z} = \mathbb{R}$) deviating from $f(x)$ less than $\epsilon$. Consequently, there could be (possibly uncountably) many candidates for $z$, implying that fine-grid search or some clever search algorithm is required. Another reason for the difficulty is because of the interaction among the (possibly high-dimensional) inputs of a (possibly nonlinear) target function. If the target function is highly nonlinear, the effect of each input (and each dimension, if multi-dimensional) in the output is intertwined in a complicated manner. Hence, it is nearly impossible to find optimal auxiliary random variables unless possible values of inputs are structured very well.

To handle such difficulty, leveraging geometric understanding of the maximal distortion, we suggest a new hypergraph-based achievable scheme that reduces the complexity of finding the best auxiliary random variables and reconstruction function. We demonstrate the hypergraph-based coding in several settings of functional compression, where it turns out to be optimal or reproduce existing achievability results without searching over the entire combinations of auxiliary random variables. Details are summarized as follows.

\begin{itemize}
	\item A hypergraph-based coding scheme is proposed, which explicitly defines auxiliary random variables.
	
	\item For functional compression with side information at decoder, it achieves the optimal rate.
	
	\item For two-terminal distributed functional compression, it achieves the sum-rate of the Berger-Tung inner bound.
	
	\item For multiple description coding for computing, it reproduces the El Gamal-Cover inner bound. Consequently, the rate-distortion region for successive refinement is achieved.
	
	\item For cascade multiple description for computing, the rate-distortion region is achieved.
	
	\item For a certain class of Markov sources, it reduces the search space of forward test channels.
\end{itemize}

\subsection{Related Literature}
Approximate computing has emerged to enable highly efficient hardware and software implementations for computing complex functions by relaxing the requirement of full-precision computation \cite{HanO2013, XuMK2016}. From the standpoint of theory, approximate computing tolerates an error up to some $\epsilon \ge 0$. For instance of point-to-point coding for computation with respect to $f$, let the perfect and approximate computations be $f(x), z(x)$, respectively. Then, the implementation of approximate computing is acceptable if $\| f(x) - z(x) \| \le \epsilon$ for all $x$. In information theory, such a notion of acceptable error was first studied by Posner and Rodemich \cite{PosnerR1971} for identity function $f(x)=x$ in the point-to-point setting, where the authors coined the term \emph{$\epsilon$-entropy} of $X$ to indicate the limit of lossy compression of $X$ up to tolerance $\epsilon$. Note that this is related to but different from the (non-stochastic) $\epsilon$-entropy \cite{KolmogorovT1959}, where the latter $\epsilon$-entropy is the logarithm of the covering number. Feizi and M\'{e}dard \cite[Thm.~43]{FeiziM2014} also briefly considered the distributed setting under maximal distortion.

Coding for computing involves compressing data efficiently so that function values for a known function $f$ can be reconstructed reliably even when the original data cannot be. When the function values are to be reconstructed losslessly, several special cases have been addressed. K\"{o}rner and Marton \cite{KornerM1979} first considered lossless distributed coding for computing for $f(x_1,x_2) = (x_1+x_2) \text{ mod } 2$ and provided the rate-distortion region for some special cases along with practical codes to achieve them. Orlitsky and Roche \cite{OrlitskyR2001} considered the setting of side information at the decoder for general $f(x_1, x_2)$ and provided a single-letter characterization based on the characteristic graph entropy. This work motivated further investigation on the characteristic graph entropy and inspired the idea of coloring graphs for network functional compression \cite{DoshiSME2010, FeiziM2014}.

For lossy coding for computing, the setting of side information at the decoder for identity function $f(x,y) = x$ was first addressed by Wyner and Ziv \cite{WynerZ1976} and for general functions was studied by Yamamoto \cite{Yamamoto1982}. However, it is not obvious how to achieve Yamamoto's rate-distortion region using practical codes as it is an existence proof using a general auxiliary random variable. Doshi et al.~\cite{DoshiSME2010} and Feizi and M\'{e}dard \cite{FeiziM2014} provided constructive achievable schemes based on characteristic graphs. However, those coding schemes have room for further improvement as we will see in Sec.~\ref{subsec:indep_source}. 

As the maximal distortion with respect to a target function $f$ is a particular instance of general distortion measures, existing achievable results for general lossy compression are immediately applicable. Specifically, we consider multiple description coding \cite{Witsenhausen1980, WolfWZ1980, WitsenhausenW1981} for which an inner bound was established by El Gamal and Cover \cite{ElGamalC1982}. It was further extended by Zhang and Berger using a common component \cite{ZhangB1987}. Multiple description coding subsumes successive refinement as a special case, for which it is shown that the El Gamal-Cover inner bound is tight \cite{EquitzC1991, Rimoldi1994}. Cascade multiple description coding for computing was also proposed, and the rate-distortion region was established by Yamamoto \cite{Yamamoto1981}. Note that the schemes used to achieve the regions in this paragraph are by no means constructive for functional compression; they need additional treatment for handling (perhaps uncountably) many candidates of approximation, which is largely unknown. 

Unlike memoryless sources for which the exact single-letter rate-distortion function is known, the lossy coding theorem for ergodic sources is known only in multi-letter form \cite{Gallager1968}. Even for a simple binary symmetric Markov source, the exact rate-distortion function is known only in a small-distortion regime \cite{Gray1970} and upper and lower bounds are known in computable form \cite{Berger1977b, JalaliW2007}. For general Markov sources, Gray \cite{Gray1971} derived a lower bound, and Jalali and Weissman \cite{JalaliW2007} derived upper and lower bounds on $R(D)$.

\subsection{Organization}
The rest of this paper is organized as follows. Sec.~\ref{sec:problem_setting} formally states definitions and problems. Sec.~\ref{sec:coding_for_computing_with_side_information} studies coding for computing with side information at the decoder under maximal distortion and discusses implications. Sec.~\ref{sec:distributed_coding} studies distributed coding for computing under maximal distortion. Secs.~\ref{sec:multiple_description_coding} and \ref{sec:cascade_mdc} respectively study multiple description coding for computing and cascade multiple description coding for computing under maximal distortion. Sec.~\ref{sec:sparse_markov} highlights the computational benefit of hypergraph-based coding for Markov sources having a sparse hypergraph. Finally, Sec.~\ref{sec:conclusion} concludes the paper.

\section{Formulation and Definition} \label{sec:problem_setting}

\subsection{Problem Formulation} \label{subsec:problem_formulation}

Consider a function of interest $f:\mathcal{X} \mapsto \mathcal{Z}$, where $\mathcal{Z}$ is a metric space with norm $\|\cdot\|$. It is acceptable in approximate computing that imprecise computations deviate from the exact ones within a given tolerance level $\epsilon \ge 0$. To model this, we consider the \emph{maximal} distortion with tolerance $\epsilon \ge 0$ defined as
\begin{align*}
	\dep(x, z) = \mathbbm{1}_{\{ \|f(x) - z\| > \epsilon \}}.
\end{align*}
We also use its length-$n$ vectorized extension $\dep(x^n, z^n) = \frac{1}{n} \sum_{t=1}^n \mathbbm{1}_{\{ \|f(x_{t}) - z_t\| > \epsilon \}}$. It is an instance of general distortion, but $\dep$ differs from the usual difference distortions (e.g., mean-squared distortion) that are widely used. To illustrate, consider the expected maximal distortion constraint,
\begin{align}
	\mathbb{E}[\dep(X, Z)] &= \mathbb{P}[\|f(X) - Z\| > \epsilon] \le D + \epsilon'(n), \label{eq:expected_max_distortion}
\end{align}
where $\epsilon'(n) \ge 0$ is a small vanishing value with the block length $n$. In this work, we only consider $D=0$, that is, when $n$ is sufficiently large, \emph{all} probable computations $(x, z)$ should meet the approximation constraint $\|f(x) - z\| \le \epsilon$ with high probability. As a contrast, consider usual difference distortion $d(x, z) = \|f(x) - z\|$ and its expected distortion constraint
\begin{align}
	\mathbb{E}[d(X, Z)] = \mathbb{E}[ \| f(X) - Z \| ] \le \epsilon. \label{eq:expected_distortion}
\end{align}
Unlike the maximal distortion, the requirement can be met even when some computations $(x, z)$ violate $\epsilon$ tolerance, i.e., $\|f(x)-z\| > \epsilon$, as long as the average distortion is less than or equal to $\epsilon$. In the circuits and systems community, computation under constraint \eqref{eq:expected_max_distortion} with $D=0$ is in particular called approximate computing, distinguishing it from related but different notions of \emph{stochastic/probabilistic computing} achieving \eqref{eq:expected_max_distortion} with $D>0$ or \eqref{eq:expected_distortion}. 
Since this work is motivated by approximate computing, constraint \eqref{eq:expected_max_distortion} with $D=0$ is considered. For side information or multiterminal problems, the single argument function $f$ can be immediately extended to a multi-argument function, e.g., $f:\mathcal{X}_1 \times \mathcal{X}_2 \mapsto \mathcal{Z}$.

Now four main problems under the maximal distortion are formally defined. For all problems, we assume that the function of interest $f$ (or multiple functions) is known to all agents. We further assume that all functions' input spaces are discrete and the output space of $f$ is a metric space with norm $\|\cdot\|$ such as $\mathbb{R}^d$ with the Euclidean distance. Also, without loss of generality, we suppose all marginal distributions of inputs are supported on the entire space. For the instance of the function $f:\mathcal{X}_1 \times \mathcal{X}_2 \mapsto \mathcal{Z}$, $p(x_i)$ is supported on $\mathcal{X}_i$. That is, no input symbol is redundant.

\subsubsection{Coding for computing with side information}

\begin{figure}[t]
	\centering
	\begin{minipage}{.45\textwidth}
		\centering
		\includegraphics[width=2.7in]{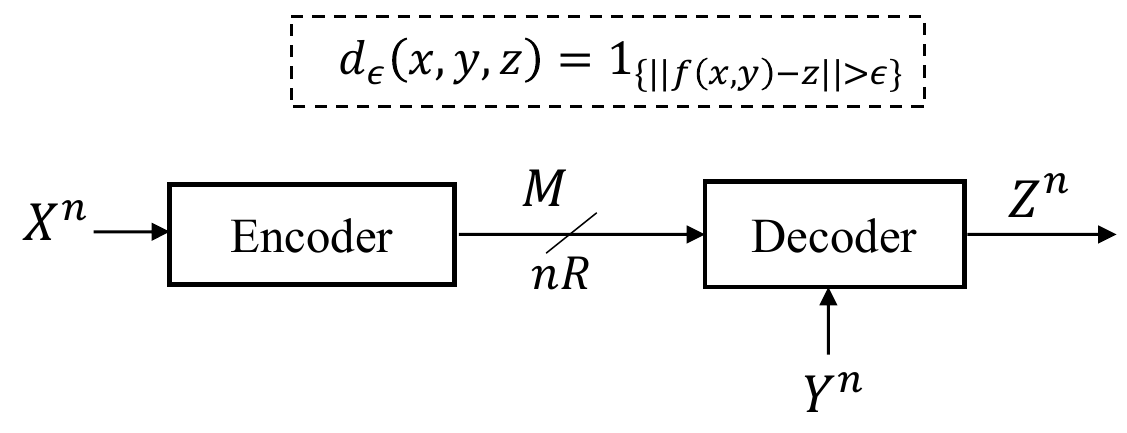}
		\caption{Coding for computing with side information at decoder in Sec.~\ref{sec:coding_for_computing_with_side_information}.}
		\label{fig:side_info}
	\end{minipage} ~~~ ~~~
	\begin{minipage}{.45\textwidth}
		\centering
		\medskip \medskip
		\includegraphics[width=2.6in]{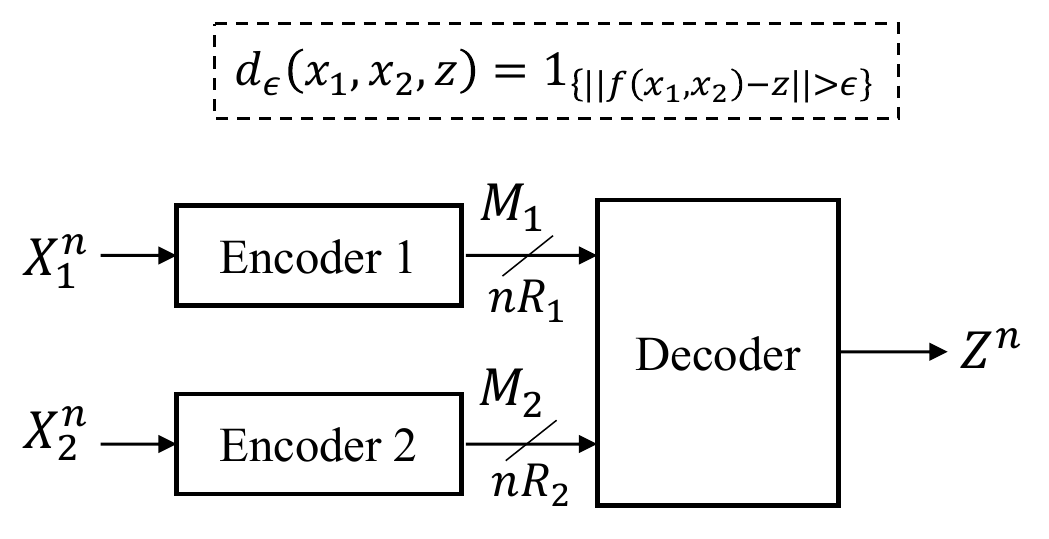}
		\caption{Distributed coding for computing in Sec.~\ref{sec:distributed_coding}.}
		\label{fig:distributed_f_compression}
	\end{minipage}
\end{figure}

Consider the problem in Fig.~\ref{fig:side_info}, where $\{(X_{t}, Y_t) \}_{t=1}^{n}$ are $n$ i.i.d.~pairs of random variables drawn from $p(x, y)$ ranging over $\mathcal{X} \times \mathcal{Y}$. The encoder observes $X^n$ and sends a message $M \in \mathcal{M} = \{1, \ldots, 2^{nR}\}$ to the decoder. The decoder receives $M$ as well as side information $Y^n$. The encoder and decoder both want the decoder to approximately compute $\{f(X_t, Y_t)\}_{t=1}^{n}$ under the maximal distortion with $\epsilon \ge 0$.

For any $R \ge 0$, we define an $(n, 2^{nR})$ code as an encoding function $\phi^{(n)}:\mathcal{X}^n \mapsto \mathcal{M}$ and decoding function $\varphi^{(n)}: \mathcal{M} \times \mathcal{Y}^n \mapsto \mathcal{Z}^n$. The expected maximal distortion of tolerance $\epsilon \ge 0$ associated with the $(n, 2^{nR})$ code is defined as
\begin{align*}
	\mathbb{E}\left[ \dep(X^n, Y^n, Z^n)  \right] &= \frac{1}{n} \sum_{t=1}^n \mathbb{E}\left[ \dep(X_{t}, Y_{t}, Z_t) \right] \\
	&= \frac{1}{n} \sum_{t=1}^n \mathbb{P}\left[ \|f(X_{t}, Y_{t}) - Z_t\| > \epsilon \right].
\end{align*}
Then, a rate $R$ is said to be \emph{achievable} if there exists a sequence of codes such that the expected maximal distortion vanishes as $n \to \infty$. The \emph{rate-distortion function} $R[\epsilon]$\footnote{We use $R[\epsilon]$ to denote the rate-distortion function to distinguish it from the standard rate-distortion function $R(D)$ since the behavior of $R[\epsilon]$ is different from that of $R(D)$. The difference will be elaborated in Sec.~\ref{subsec:ptp_compression}. For the same reason, we use $\mathcal{R}[\cdot]$ for rate-distortion or achievable regions instead of $\mathcal{R}(\cdot)$.} is the infimum of all achievable rates.

\subsubsection{Distributed coding for computing}

Consider the problem shown in Fig.~\ref{fig:distributed_f_compression}, where $\{(X_{1t},X_{2t})\}_{t=1}^{n}$ are $n$ i.i.d.~pairs of random variables from $p(x_1, x_2)$ ranging over $\mathcal{X}_1 \times \mathcal{X}_2$. The encoder $i$ observes $X_i^n$ and sends a message $M_i \in \mathcal{M}_i := \{1, \ldots, 2^{nR_i}\}$ to the decoder so that the decoder can reconstruct $\{f(X_{1t}, X_{2t})\}_{t=1}^n$ approximately under the maximal distortion with tolerance $\epsilon \ge 0$. 

For any $(R_1, R_2) \in \mathbb{R}_+^2$, we define an $(n, 2^{nR_1}, 2^{nR_2})$ code as encoding functions $\phi_i^{(n)}:\mathcal{X}_i^n \mapsto \mathcal{M}_i$, $i \in \{1,2\}$ and a decoding function $\varphi^{(n)}: \mathcal{M}_1 \times \mathcal{M}_2 \mapsto \mathcal{Z}^n$. The expected maximal distortion of tolerance $\epsilon$ associated with the $(n, 2^{nR_1}, 2^{nR_2})$ code is defined as
\begin{align*}
	\mathbb{E}\left[ \dep(X_1^n, X_2^n, Z^n)  \right] = \frac{1}{n} \sum_{t=1}^n \mathbb{P}\left[ \|f(X_{1t}, X_{2t}) - Z_t\| > \epsilon \right].
\end{align*}
Then, a rate pair $(R_1, R_2)$ is said to be \emph{achievable} if there exists a sequence of codes such that the expected maximal distortion vanishes as $n \to \infty$. The \emph{rate-distortion region} $\mathcal{R}[\epsilon]$ is the closure of the set of all achievable rate pairs.

\subsubsection{Multiple description coding for computing}
\begin{figure}[t]
	\centering
	\begin{minipage}{.45\textwidth}
		\centering
		\includegraphics[width=2.7in]{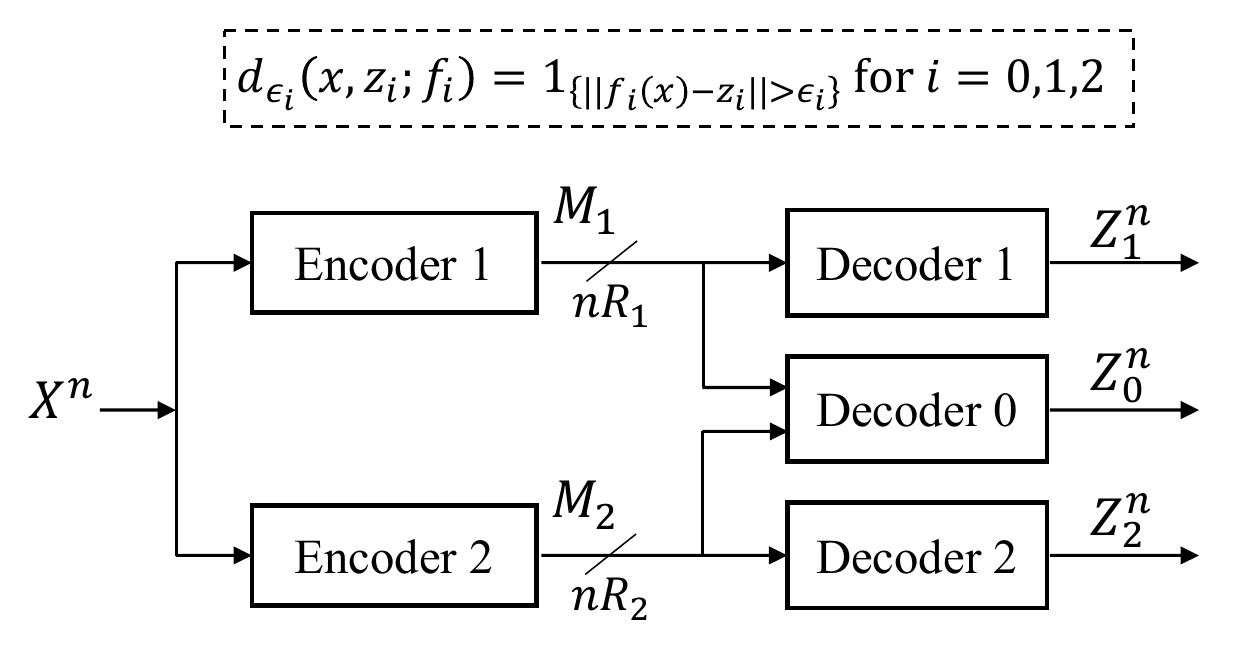}
		\caption{Multiple description coding for computing in Sec.~\ref{sec:multiple_description_coding}.}
		\label{fig:mdc}
	\end{minipage} ~~~ ~~~
	\begin{minipage}{.45\textwidth}
		\centering
		\bigskip \medskip
		\includegraphics[width=3.1in]{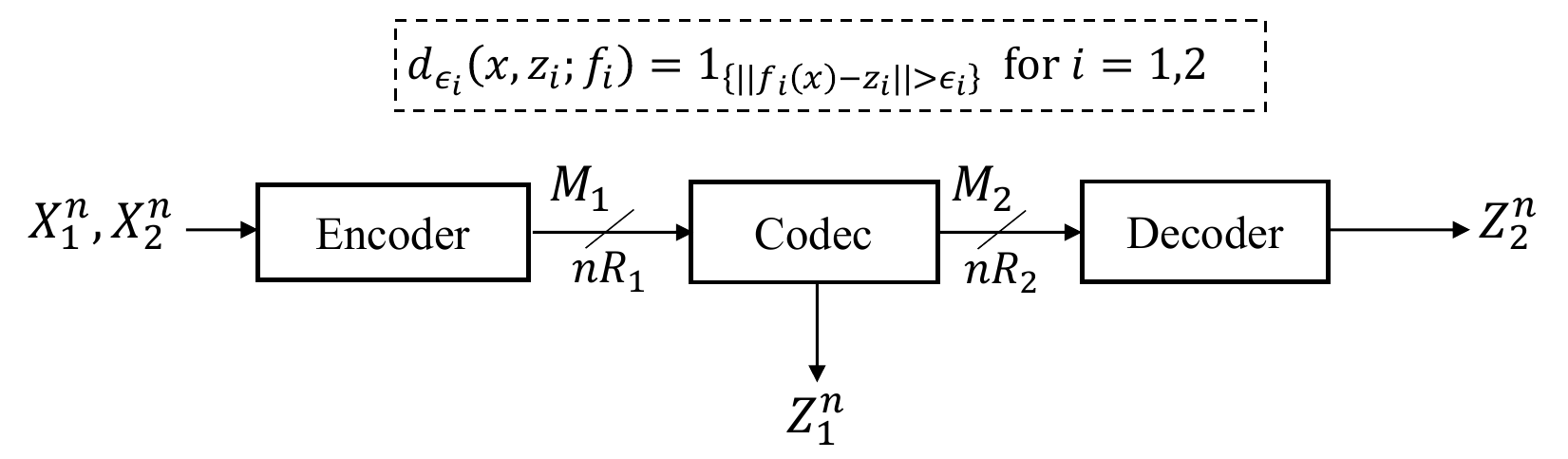}
		\caption{Cascade multiple description coding for computing in Sec.~\ref{sec:cascade_mdc}.}
		\label{fig:cascade_mdc}
	\end{minipage}
\end{figure}

Consider the problem in Fig.~\ref{fig:mdc}, where $\{X_t\}_{t=1}^n$ are $n$ i.i.d.~random variables from $p(x)$ over $\mathcal{X}$. There are three functions $f_i:\mathcal{X} \mapsto \mathcal{Z}_i, i \in \{0,1,2\}$ known to all encoders and decoders. Encoder $1$ and $2$ both observe $\{X_t\}_{t=1}^n$, and encoder $i$ sends a message $M_i \in \mathcal{M}_i = \{1, 2, \ldots, 2^{nR_i}\}$ to decoder $i$ as well as to common decoder $0$. On receiving the messages, decoder $i\in\{0,1,2\}$ reconstructs $\{Z_{it}\}_{t=1}^{n}$ so that $\{f_i(X_t)\}_{t=1}^n$ can be recovered approximately under the maximal distortion for $f_i$ with tolerance $\epsilon_i$.

For any $(R_1, R_2) \in \mathbb{R}_+^2$, we define an $(n, 2^{nR_1}, 2^{nR_2})$ code as two encoding functions $\phi_i^{(n)}:\mathcal{X}_i^n \mapsto \mathcal{M}_i, i \in \{1,2\}$ and three decoding functions $\varphi_i^{(n)}: \mathcal{M}_i \mapsto \mathcal{Z}_i^n, i \in \{1,2\}$ and $\varphi_0^{(n)}: \mathcal{M}_1 \times \mathcal{M}_2 \mapsto \mathcal{Z}_0^n$. The three expected maximal distortion constraints of tolerance $\epsilon_i$ each associated with the $(n, 2^{nR_1}, 2^{nR_2})$ code are defined as $\mathbb{E}\left[ d_{\epsilon_i}(X^n, Z_i^n;f_i) \right] = \frac{1}{n} \sum_{t=1}^n \mathbb{P}\left[ \|f_i(X_{t}) - Z_{it}\| > \epsilon_i \right], ~~~ i \in \{0,1,2\}$. Then, a rate pair $(R_1, R_2) \in \mathbb{R}_+^2$ is said to be \emph{achievable} if there exists a sequence of codes such that all three expected maximal distortions vanish as $n \to \infty$. The \emph{rate-distortion region} $\mathcal{R}[\epsilon_0, \epsilon_1, \epsilon_2]$ is the closure of the set of all achievable rate pairs. In addition, when decoder $2$ does not exist, the problem setting is called the successive refinement problem.

\subsubsection{Cascade multiple description coding for computing}
Consider the problem in Figure \ref{fig:cascade_mdc}, where $\{X_{1t}, X_{2t}\}_{t=1}^n$ are $n$ i.i.d.~pairs of random variables from $p(x_1, x_2)$ over $\mathcal{X}_1 \times \mathcal{X}_2$. There are two functions $f_i:\mathcal{X}_i \mapsto \mathcal{Z}_i, i \in \{1,2\}$ of interest. The encoder first observes $X_1^n, X_2^n$ and sends a message $M_1 \in \mathcal{M}_1 = \{1, \ldots, 2^{nR_1}\}$ to the codec. On receiving $M_1$, the codec reconstructs $\{f_1(X_{1t})\}_{t=1}^{n}$ approximately under the maximal distortion of tolerance $\epsilon_1$. Then, the codec generates another message $M_2 \in \mathcal{M}_2 = \{1, 2, \ldots, 2^{nR_2}\}$ and sends to the decoder. The decoder reconstructs $\{f_2(X_{2t})\}_{t=1}^{n}$ approximately under the maximal distortion of tolerance $\epsilon_2$.

For any $(R_1, R_2) \in \mathbb{R}_+^2$, we define an $(n, 2^{nR_1}, 2^{nR_2})$ code as two encoding functions $\phi_1^{(n)}:\mathcal{X}_1^n \times \mathcal{X}_2^n \mapsto \mathcal{M}_1$ and $\phi_2^{(n)}:\mathcal{M}_1 \mapsto \mathcal{M}_2$ and two decoding functions $\varphi_i^{(n)}: \mathcal{M}_i \mapsto \mathcal{Z}_i^n, i \in \{1,2 \}$. The two expected maximal distortion constraints of tolerance $\epsilon_i$ each associated with the $(n, 2^{nR_1}, 2^{nR_2})$ code are defined as $\mathbb{E}\left[ d_{\epsilon_i}(X_i^n, Z_i^n;f_i) \right] = \frac{1}{n} \sum_{t=1}^n \mathbb{P}\left[ \|f_i(X_{it}) - Z_{it}\| > \epsilon_i \right]$ for $i \in \{1,2\}$. Then, a rate pair $(R_1, R_2)$ is said to be \emph{achievable} if there exists a sequence of codes such that the two expected maximal distortions both vanish as $n \to \infty$. The \emph{rate-distortion region} $\mathcal{R}[\epsilon_1, \epsilon_2]$ is the closure of the set of all achievable rate pairs.

\subsection{Definitions} \label{subsec:def}
Here, we describe the construction of maximal $\epsilon$-characteristic hypergraphs that will be used in sequel. A hypergraph $G$ is a pair $G = (\mathcal{X}, \mathcal{E})$, where $\mathcal{X}$ is a set of vertices and $\mathcal{E}$ is a set of hyperedges, i.e., non-empty subsets of $\mathcal{X}$. Formally, $\mathcal{E} \subseteq 2^\mathcal{X} \setminus \emptyset$, where $2^\mathcal{X}$ is the powerset of $\mathcal{X}$ \cite{Bretto2013}. We further define two notions; $\epsilon$-characteristic hypergraphs for a function $f$ and the maximal set of hyperedges.

\begin{defi} \label{def:smallest_enclosing_circle}
	For a set of $k$ points $\{z_1, \ldots, z_k\}$, $z_i \in \mathcal{Z}$, where $\mathcal{Z}$ is a metric space, the \emph{smallest enclosing circle (SEC)} is a circle (sphere) that contains all of the $k$ points with the smallest radius. That is, the SEC is given by a center $c$ and radius $r$ such that $\|z_i - c\| \leq r$.
\end{defi}

\begin{defi} \label{def:ep_approx}
	A function $g:\mathcal{X} \mapsto \mathcal{Z}$, where $\mathcal{Z}$ is a metric space, is a \emph{pointwise $\epsilon$-approximation} to $f:\mathcal{X} \mapsto \mathcal{Z}$ at $x \in \mathcal{X}$ if $\|f(x) - g(x)\| \leq \epsilon$. If $g$ is a pointwise $\epsilon$-approximation to $f$ at every $x \in \mathcal{X}$, it is an \emph{$\epsilon$-approximation} to $f$.
\end{defi}

Note that the input of $f$ can be general; in source coding with side information, it is a Cartesian product of source and side information symbols, i.e., $\mathcal{X} \times \mathcal{Y}$.
\begin{defi} \label{def:e_characteristic_hypergraph}
	An \emph{$\epsilon$-characteristic hypergraph} for $f:\mathcal{X} \times \mathcal{Y} \mapsto \mathcal{Z}$ is a hypergraph $G^{\epsilon, f} = (\mathcal{X}, \mathcal{E}^{\epsilon, f})$ that satisfies the following two properties.
	\begin{itemize}
		\item (SEC property) For any $w \in \mathcal{E}^{\epsilon, f}$, the radius of the SEC of points $\{f(x,y): x \in w \text{ and } p(x,y) > 0 \}$ is less than or equal to $\epsilon$ for all $y$.
		
		\item (Edge cover) Every $x \in \mathcal{X}$ belongs to at least one $w \in \mathcal{E}^{\epsilon, f}$.
	\end{itemize}
\end{defi}
Informally speaking, a hyperedge is a collection of inputs that produce similar function values within $\epsilon$ for all $y$. In other words, an $\epsilon$-characteristic hypergraph $G^{\epsilon, f} = (\mathcal{X}, \mathcal{E}^{\epsilon, f})$ can give us an $\epsilon$-approximation to $f(x, y)$ at $(x,y)$ such that $x \in w \in \mathcal{E}$ and $p(x,y) > 0$. Also, note that if $\epsilon=0$, then the $\epsilon$-characteristic hypergraph essentially reduces to the characteristic graph in \cite{Witsenhausen1976, OrlitskyR2001}. Therefore, it can be thought of as a generalization of the characteristic graph.

Let $\Gamma^{\epsilon, f} = \cup \mathcal{E}^{\epsilon, f}$ be the union of sets of hyperedges of all $\epsilon$-characteristic hypergraphs. Then, we can see two properties that: 1) $\Gamma^{0, f}$ is nonempty as the set of all singletons satisfies the SEC and edge cover properties, and 2) $\Gamma^{\epsilon_1, f} \subseteq \Gamma^{\epsilon_2, f}$ for $0 \leq \epsilon_1 \leq \epsilon_2$.  As will be shown below by examples, there are multiple hyperedge sets satisfying the SEC and edge cover properties, and therefore, multiple $\epsilon$-characteristic hypergraphs exist in general. Among them, it is sufficient to consider a particular hypergraph, which we call the maximal $\epsilon$-characteristic hypergraph, without performance degradation in distortion and rate. Its hyperedge set is referred to as the maximal set of hyperedges or maximal hyperedge set.
\begin{defi} \label{def:max_characteristic_graph}
	The union of hyperedge sets of all $\epsilon$-characteristic hypergraphs is denoted by $\Gamma^{\epsilon, f} = \cup \mathcal{E}^{\epsilon, f}$. Then, the maximal set of hyperedges $\Gamma_m^{\epsilon, f}$ is obtained by deleting hyperedges that are a subset of other hyperedges in $\Gamma^{\epsilon, f}$. Edges in $\Gamma_{m}^{\epsilon, f}$ are called maximal. A hypergraph $G_m^{\epsilon, f} = (\mathcal{X}, \Gamma_m^{\epsilon, f})$ is the maximal $\epsilon$-characteristic hypergraph for $f$.
\end{defi}
Note that $\Gamma_m^{\epsilon,f}$ is uniquely determined by $\epsilon, f$, and underlying probabilities. In general, $\Gamma_m^{\epsilon,f}$ is constructed by a brute-force method, but it can be determined more simply depending on the source and $f$, see Example \ref{ex:birth_death}. The following examples illustrate $\epsilon$-characteristic hypergraphs and maximal sets of hyperedges for functional compression with side information.

\begin{ex} \label{ex:hypergraph1}
	Let $X$ be a uniform random variable over $\{1,2,3\}$, $Y \equiv 0$, and $\epsilon=0$. The function of interest is $f(x,y) = x$. As $|\mathcal{X}|=3$, there are $2^3-1$ possible hyperedges, but only singleton sets satisfy the SEC condition for $\epsilon=0$. Note that the set of all singletons is an edge cover. Therefore, the $0$-characteristic hypergraph is $G^{0, f} = (\mathcal{X}, \mathcal{E}^{0, f})$ with $\mathcal{E}^{0, f} =\{ \{1\}, \{2\}, \{3\} \}$, which is maximal as well, i.e., $\Gamma_m^{0, f} = \mathcal{E}^{0, f}$ and $G_m^{0,f} = G^{0,f}$.
\end{ex}
\begin{ex} \label{ex:hypergraph2}
	Consider Example \ref{ex:hypergraph1} again, but with $p_X(1)=p_X(3)=\frac{1}{2}, p_X(2)=0$.\footnote{It violates our assumption that all marginal probabilities are positive, but we include it for expository purposes. Extending this argument to when $Y$ varies, we can see that if $p(x,y')=0$ for some $y'$, then $\{x\}$ can be added to an arbitrary hyperedge to reduce information rate, as long as the addition does not violate the SEC property at another $y'' \ne y'$.} In this case, hyperedges satisfying the SEC conditions are $\{1\}, \{2\}, \{3\}, \{1,2\}$, and $\{2,3\}$, from which we can obtain four hyperedge sets that are edge covers and have no proper subset in each: $\mathcal{E}_1 = \{ \{1\}, \{2\}, \{3\} \}$, $\mathcal{E}_2 = \{ \{1\}, \{2,3\} \}$, $\mathcal{E}_3 = \{ \{1,2\}, \{3\} \}$, and $\mathcal{E}_4 = \{ \{1,2\}, \{2,3\} \}$. Note that they construct four $0$-characteristic hypergraphs. To obtain the maximal hyperedge set, taking union $\cup_i \mathcal{E}_i = \{\{1\}, \{2\}, \{3\}, \{1,2\}, \{2,3\} \}$ and then deleting elements that are a subset of others give the maximal set of hyperedges  $\{\{1,2\}, \{2,3\} \}$. Therefore, $G_m^{0, f} = (\mathcal{X}, \Gamma_m^{\epsilon, f})$ with $\Gamma_m^{\epsilon, f} = \mathcal{E}_4$.
\end{ex}

\begin{ex} \label{ex:hypergraph3}
	Consider Example \ref{ex:hypergraph1} again, but with $\epsilon = 1$. In this case, hyperedges satisfying the SEC conditions are $\{1\}$, $\{2\}$, $\{3\}$, $\{1,2\}$, and $\{2,3\}$, from which we can obtain the same four hypergraphs as in Example \ref{ex:hypergraph2} that are $1$-characteristic hypergraphs having no proper subset in hyperedge sets. In this case, $G_m^{1, f} = (\mathcal{X}, \Gamma_m^{\epsilon, f})$ with $\Gamma_m^{\epsilon, f} = \{ \{1,2\}, \{2,3\} \}$.
\end{ex}

We can extend the maximal $\epsilon$-characteristic hypergraph to that for multiterminal functional compression problems as follows. This is similar to but different from the graph construction of \cite{FeiziM2014} in capturing interaction with another source. This will be detailed in Sec.~\ref{subsec:indep_source}.

\begin{defi} \label{def:epsilon_hypergraph_pair}
	Let $f: \mathcal{X}_1\times \mathcal{X}_2 \mapsto \mathcal{Z}$ be a function where $\mathcal{Z}$ is a metric space. Also, let $G_i^{\epsilon, f} = (\mathcal{X}_i, \mathcal{E}_i)$, $i \in \{1,2\}$. Then, a pair of hypergraphs $(G_1^{\epsilon, f}, G_2^{\epsilon, f})$ is \emph{an $\epsilon$-characteristic hypergraph pair} with respect to $f$ if the following two properties are satisfied.
	\begin{itemize}
		\item (SEC property) For any $w_1 \in \mathcal{E}_1$ and $w_2 \in \mathcal{E}_2$ the radius of the SEC of the set of points $\{f(x_1,x_2): x_1 \in w_1, x_2 \in w_2,\text{ and } p(x_1,x_2) > 0\}$ is less than or equal to $\epsilon$.
		
		\item (Edge cover) Every $x_i \in \mathcal{X}_i$ belongs to at least one $w_i \in \mathcal{E}_i^{\epsilon, f}$ for $i \in \{1,2\}$
	\end{itemize}
\end{defi}

\begin{defi} \label{def:max_pair_hypergraph}
	For each $i$, the union of hyperedge sets of all $\epsilon$-characteristic hypergraph $G_i^{\epsilon, f}=(\mathcal{X}, \mathcal{E}_i^{\epsilon, f})$ is denoted by $\Gamma_i^{\epsilon, f} = \cup \mathcal{E}_i^{\epsilon, f}$. Then, the pair of maximal hyperedge sets $(\Gamma_{m, 1}^{\epsilon, f}, \Gamma_{m, 2}^{\epsilon, f})$ is obtained by deleting hyperedges $w \in \Gamma_i^{\epsilon, f}$ that are subsets of other hyperedges in $\Gamma_i^{\epsilon, f}$. A hypergraph pair $(G_{m, 1}^{\epsilon, f}, G_{m, 2}^{\epsilon, f})$, where $G_{m, i}^{\epsilon, f} = (\mathcal{X}, \Gamma_{m, i}^{\epsilon, f})$, is the pair of maximal $\epsilon$-characteristic hypergraphs for $f$.
\end{defi}

\section{Coding for Computing with Side Information} \label{sec:coding_for_computing_with_side_information}

\subsection{Point-to-point Functional Compression} \label{subsec:ptp_compression}
Let us consider the simplest point-to-point functional compression problem without side information to illustrate the challenges in existing literature. Suppose that $X$ is a source and $f:\mathcal{X} \mapsto \mathcal{Z}$ is a function of interest. Then, we wish to compress $X$ so that a decoder can approximate $f(X)$ under maximal distortion. This is a special case of coding for computing with side information since the function of interest $g(X,Y)$ is indeed $f(X)$ for some $f$ if $Y \equiv \text{constant}$. Recall that $\dep(x, z) = \dep(x, z; f) = \mathbbm{1}_{ \{ \| f(x) - z \| > \epsilon \} }$ is an instance of general distortion in rate-distortion theory. Hence the optimal rate is
\begin{align}
	R[\epsilon] = \min_{p(z|x): \mathbb{E}\left[ \dep(X, Z) \right] = 0} I(X;Z). \label{eq:functional_compression}
\end{align}

Recall the standard source coding: $\min I(X;\hat{X})$ where the minimization is over $p(\hat{x}|x)$ satisfying a distortion constraint. The reconstruction space $\hat{\mathcal{X}}$ is usually finite and predetermined. Thus, finding a test channel $p(\hat{x}|x)$ can be performed via optimization algorithms such as Blahut-Arimoto. However, in \eqref{eq:functional_compression}, the space for $Z$ is not fixed and could even be an uncountably large subset of the entire $\mathcal{Z}$, e.g., an interval in $\mathbb{R}$, since $\epsilon$ deviation is allowed. Then, to find the optimal $p(z|x)$, an optimization algorithm must search $p(z|x)$ over an infinite dimensional space (e.g., a probability density over an interval) or over a very large dimensional space even after discretization. We propose hypergraph-based coding that successfully solves the challenge. Note that the number of hyperedges is finite.
\begin{thm} \label{thm:warm_up}
	Let $G_m^{\epsilon, f} = (\mathcal{X}, \Gamma_m^{\epsilon, f})$ be the maximal $\epsilon$-characteristic hypergraph for $f$. Then,
	\begin{align*}
		R[\epsilon] &= \min_{p(z|x): \mathbb{E}\left[ \dep(X, Z) \right] = 0} I(X;Z) = \min_{p(w|x): X \in W \in \Gamma_m^{\epsilon, f}} I(X;W),
	\end{align*}
	where ``$p(w|x): X \in W \in \Gamma_m^{\epsilon, f}$'' indicates $p(w|x)$ can be nonzero only when $x \in w \in \Gamma_{m}^{\epsilon, f}$.
\end{thm}
\begin{IEEEproof}[Proof Sketch]
	After constructing $G_m^{\epsilon,f}$, the encoder encodes hyperedges instead of the source. The decoder reconstructs the center of function values corresponding to the hyperedges. Details are omitted since this is a special case of Thm.~\ref{thm:coding_for_computing_side_info} when $Y =  \text{const}$.
\end{IEEEproof}

The theorem implies that instead of optimizing over the entire space $\mathcal{Z}$ that is perhaps continuous, it is sufficient to optimize over the maximal set of hyperedges that is finite and uniquely determined by $\epsilon, f$, and underlying probability. Moreover, declaring the center of the SEC of points $\{f(x): x \in w \in \Gamma_m^{\epsilon, f} \}$ achieves the maximal distortion of tolerance $\epsilon$ as we desired. Therefore, Thm.~\ref{thm:warm_up} greatly reduces the complexity of optimizing $Z$, i.e., it is constructive.

Before proceeding to our main problems, several remarks should be noted. First, unlike a standard rate-distortion curve, $R[\epsilon]$ is pieceswise constant in $\epsilon$. To depict $R[\epsilon]$, consider Example \ref{ex:discontinuous} below, a generalization of Example \ref{ex:hypergraph1} with general $\epsilon$. As it is different from that of standard rate-distortion function $R(D)$, we use $R[\epsilon]$ notation.
\begin{figure}[t]
	\centering
	\includegraphics[width=3.0in]{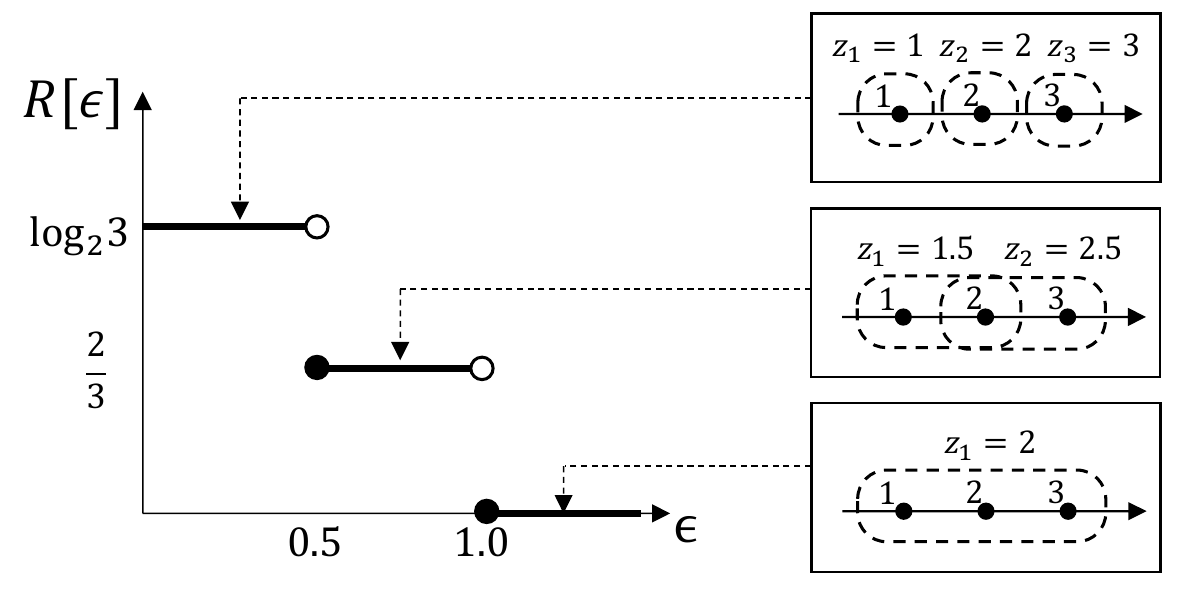}
	\caption{$R[\epsilon]$ for Example \ref{ex:discontinuous}. The maximal set of hyperedges and corresponding reconstruction function are also shown.}	\label{fig:discontinuous_example}
\end{figure}
\begin{ex} \label{ex:discontinuous}
	Let $X$ be a uniform random variable over $\{1,2,3\}$ and the function of interest is $f(x) = x$. Then, we can derive that
	\begin{align*}
		R[\epsilon] = \begin{cases}
			\log_2 3 & \text{if }~ \epsilon < 0.5, \\
			\frac{2}{3} & \text{if }~ 0.5 \le \epsilon < 1, \\
			0 & \text{if }~ \epsilon \ge 1.
		\end{cases}
	\end{align*}
	The maximal set of hyperedges together with corresponding approximate function is in Fig.~\ref{fig:discontinuous_example}.
\end{ex}
The reason for the discontinuity is that standard curve-convexifying technique is inapplicable. Recall the difference between \eqref{eq:expected_max_distortion} with $D=0$ (approximate computing) and \eqref{eq:expected_distortion} (stochastic computing). Unlike stochastic computing, which controls the average distortion by varying $D$ with fixed $d$, approximate computing parameterizes the distortion measure by $\epsilon$ and controls the amount of deviation by varying $\epsilon$. Expected distortion is still required to be zero under $\dep$. Hence, as varying $\epsilon$, approximate computing indeed solves problems with different distortion measures. This difference prevents time-sharing across multiple $\epsilon$ values from convexifying the rate-distortion curve under the maximal distortion. Note that the maximal distortion aims at computing \emph{all} values of $\{f(x)\}$ within $\epsilon$ tolerance---thus any convex combination between two points $p_1(w|x)$ over $\Gamma_m^{\epsilon_1, f}$ and $p_2(w|x)$ over $\Gamma_m^{\epsilon_2, f}$ such that $\epsilon_2 > \epsilon_1$ will achieve $\epsilon_2$ tolerance at best. Also, to compress the source, we use hyperedges and SECs of $\{f(x): x \in w \}$ with a radius no greater than $\epsilon$. This in turn implies that there are at most a finite number of hyperedges and SECs since the number of points $\{f(x)\}$ to approximate is finite. Moreover, by the SEC law, the maximal hyperedge set remains unchanged for some interval $\epsilon$, which in turn leads to the flat behavior of achievable rates.

It is also remarkable that as long as $p(x) > 0$ for all $x$, the construction of the maximal $\epsilon$-characteristic hypergraph is independent of $p(x)$. In other words, less likely symbols and more likely symbols are treated equally in the maximal hypergraph construction. This is because the goal is to reconstruct \emph{all} $f(x)$ within tolerance $\epsilon$ regardless of its probability. In this sense, the construction of the maximal hypergraph is universal over $p(x)$ as long as all $x$ have a nonzero probability. This generalizes to multiterminal functional compression scenarios in the sequel that the maximal hypergraph is universal over different underlying probabilities as long as all source outcomes are possible.

\subsection{Coding for Computing with Side Information at Decoder} \label{subsec:optimal_rate}
Consider the first main problem, coding for computing with side information at the decoder. Since the maximal distortion is an instance of a general distortion measure, the existing optimal rate expression by \cite{Yamamoto1982, WynerZ1976} still holds.
\begin{thm}[\cite{Yamamoto1982, WynerZ1976}]
	Consider $d:\mathcal{X} \times \mathcal{Y} \times \mathcal{Z} \mapsto \mathbb{R}$. Then, the smallest rate $R(D;d)$ allowing the expected distortion up to $D \ge 0$ is
	\begin{align*}
		R(D;d) = \min_{\substack{V-X-Y \\ \exists g \text{ s.t. } \mathbb{E}[d(X, Y, g(V, Y))] \le D }} I(X;V|Y).
	\end{align*}
\end{thm}

According to the theorem, the rate-distortion function for the maximal distortion is $R[\epsilon] = R(0;\dep)$ since the maximal distortion is indeed $\dep(x, y, z) = \mathbbm{1}_{ \{ \| f(x,y) - z \| > \epsilon \} }$ with $D=0$. However, it is in general challenging to construct an auxiliary random variable $V$ achieving $R(0;\dep)$ as previously stated. The following theorem shows that constructing $V$ based on the maximal $\epsilon$-characteristic hypergraph is indeed optimal.

\begin{thm} \label{thm:coding_for_computing_side_info}
	Let $G_m^{\epsilon, f} = (\mathcal{X}, \Gamma_m^{\epsilon, f})$ be the maximal $\epsilon$-characteristic hypergraph with respect to $f$. Then,
	\begin{align*}
		R[\epsilon] = \min_{\substack{W-X-Y \\ X \in W \in \Gamma_m^{\epsilon, f} }} I(X;W|Y),
	\end{align*}
	where ``$X \in W \in \Gamma_m^{\epsilon, f}$'' indicates $p(w|x)$ can be nonzero only when $x \in w \in \Gamma_{m}^{\epsilon, f}$.
\end{thm}
\begin{IEEEproof}
	Since $R[\epsilon] = R(0;\dep)$, it is sufficient to show that
	\begin{align*}
		\min_{\substack{V-X-Y \\ \exists g \text{ s.t. } \mathbb{E}[\dep(X, Y, g(V, Y))] =0 }} I(X;V|Y) = \min_{\substack{W-X-Y \\ X \in W \in \Gamma_m^{\epsilon, f} }} I(X;W|Y)
	\end{align*}
	
	Let us first prove ``$\le$'' direction. Fix $W$ that achieves the minimum of the right side. Then, by the definition of $\epsilon$-characteristic hypergraph, for each $w \in \Gamma_m^{\epsilon, f}$, we can find the center of the SEC that $\epsilon$-approximates $f(x, y)$ at every $x \in w$ and $y \in \mathcal{Y}$ such that $p(x, y) > 0$. Let $\widetilde{g}: \Gamma_m^{\epsilon, f} \times \mathcal{Y} \mapsto \mathcal{Z}$ be the function that finds such centers. Then, $\widetilde{g}$ is an $\epsilon$-approximation to $f(x, y)$ whenever $p(w, x, y) > 0$. Since $W$ and $\widetilde{g}$ satisfy the condition of the minimum of the left side, it is a particular instance of $V$ and $g$. In addition, as $V-W-X-Y$ holds, we have $I(X;V|Y) \le I(X;W|Y)$ by the data processing inequality. Taking minimum over other $V$s proves ``$\le$'' direction.
	
	Next, we show ``$\ge$'' direction. Fix $V$ and $g$ that achieve the minimum of the left side. We will construct a particular instance of $\epsilon$-characteristic hypergraph $G^{\epsilon, f} = (\mathcal{X}, \mathcal{E}^{\epsilon, f})$ from $V$ and show that $I(X;V|Y) \ge I(X;W|Y)$ for possibly non-maximal $\epsilon$-characteristic hypergraph. Then, considering test channels over the maximal hyperedges $\breve{W}$ further reduces the rate, which concludes ``$\ge$'' direction.
	
	Let $p(v,x,y)$ be the probability distribution over $(V, X, Y)$ and $\widetilde{w}(v)$ be the inverse image of $v$ such that $\widetilde{w}(v) = \{x: p(v,x) > 0\}$. Also, define a Markov chain $W-V-(X,Y)$ such that 
	\begin{align} \label{eqn:w_def}
		p(w|v,x,y) = 
		\begin{cases}
			1 & \text{if } w = \widetilde{w}(v), \\
			0 & \text{otherwise.}
		\end{cases}
	\end{align}
	Here, the sample space of the random variable $W$ is the set of subsets of $\mathcal{X}$. Note that $p(w,x)>0$ implies that there is a $v$ such that $p(v,x) > 0$ and $w = \hat{w}(v)$. Then, by construction of $\widetilde{w}$ and \eqref{eqn:w_def}, $x \in w$ whenever $p(w,x) > 0$. Consider the case $p(w)>0$. Moreover, if $p(x,y)>0$ and $x\in w$, then the radius of the SEC of $\{f(x,y): x \in w \text{ and } p(x,y) > 0\}$ is less than or equal to $\epsilon$. This is because $V-X-Y$ forms a Markov chain, and $\{p(x, y)>0, p(v, x)>0\}$ implies $p(v, x, y)>0$. Now, to ensure that $\mathbb{E}[\dep(X, Y, g(V, Y))] =0$, we must have that the radius of the SEC of $\{f(x,y): x \in w \text{ and } p(x,y) > 0\}$ is less than or equal to $\epsilon$. We can then obtain a series of Markov chains and conclude that $I(X;W|Y) \leq I(X;V|Y)$ holds for $G^{\epsilon, f}$. Detailed proof of the last part can be found in App.~\ref{app:pf_cfc_side_info}.

	Note that the argument holds for the $\epsilon$-characteristic hypergraph derived from $V$ and $g$, possibly non-maximal. However, optimizing over the maximal $\epsilon$-characteristic hypergraph further minimizes the rate without affecting the distortion. To see this, let $\mathcal{E}^{\epsilon, f}$ be a non-maximal set of hyperedges of an $\epsilon$-characteristic hypergraph. Since $\mathcal{E}^{\epsilon, f}$ is non-maximal, we can (perhaps stochastically) map an edge in $\mathcal{E}^{\epsilon, f}$ to a bigger maximal edge in $\Gamma_m^{\epsilon, f}$ that includes the non-maximal one. That is, $w$ can be further mapped into maximal $\breve{w} \in \Gamma_m^{\epsilon, f}$ satisfying $\breve{W}-W-X-Y$ without increase in distortion. Then, the data processing inequality gives
	\begin{align*}
		\min_{\substack{W-X-Y \\ X \in W \in \mathcal{E}^{\epsilon, f} }} I(X;W|Y) \ge I(X;\breve{W}|Y).
	\end{align*}
	Note that the test channels $p(\breve{w}|x)$ obtained via the Markov chain are only a part of all possible $p(\breve{w}|x)$ due to the Markov restriction. Hence, taking test channels directly over $\Gamma_{m}^{\epsilon, f}$ further minimizes the rate, i.e.,
	\begin{align*}
		\min_{\substack{W-X-Y \\ X \in W \in \mathcal{E}^{\epsilon, f} }} I(X;W|Y) \ge \min_{\substack{\breve{W}-X-Y \\ X \in \breve{W} \in \Gamma_m^{\epsilon, f} }} I(X;\breve{W}|Y).
	\end{align*}
\end{IEEEproof}

Note that setting $\epsilon = 0$ reproduces the result in \cite[Thm.~2]{OrlitskyR2001} for lossless functional compression with side information.

\subsection{Approximation-compression Separation when $\epsilon=0$}

In the above, we have discussed that the optimal rate can be achieved by the hypergraph-based coding scheme. The rate is in the form of optimizing $I(X;W|Y)$ over the probability $p(x,w)$ over the maximal set of hypergraphs $\Gamma_m^{\epsilon, f}$. As the reconstruction alphabet is fixed to be the maximal hyperedge set, this can be solved by numerical algorithms such as Blahut-Arimoto, but one may wonder if we can further simplify the optimization process. We provide an affirmative answer for the $\epsilon=0$ case.

Note that $I(X;W|Y) = H(W|Y) - H(W|X,Y) = H(W|Y) - H(W|X)$ as $W-X-Y$. Therefore, when $H(W|X)=0$, i.e., each $x$ fully determines $w$ that it belongs to, or in other words, each $x$ belongs to exactly one hyperedge, the rate reduces to $H(W|Y)$. This in turn implies $p(w|x) = \mathbbm{1}_{\{ x \in w \}}$, which gives $p(w|y)$ in a closed form. Therefore, the optimal coding can be simply achieved by (approximation) finding the maximal $\epsilon$-characteristic hypergraph and then (compression) compressing $w \in \mathcal{E}$ according to Slepian-Wolf with two sources $W$ and $Y$ as if $Y$ is compressed at its full rate $H(Y)$. The modular design achieves the optimal rate without any rate loss as long as $H(W|X)=0$. To this end, we investigate a condition under which hyperedges in the maximal hyperedge set do not overlap. That is, $H(W|X)=0$.

\begin{condition} \label{cond:1}
	For every $y \in \mathcal{Y}$, if $f(x_1,y) \neq f(x_2,y)$ for some $x_1, x_2 \in \mathcal{X}$, then either $p(x_1,y) = p(x_2,y) = 0$ or $p(x_1,y)>0, p(x_2,y)>0$.
\end{condition}

Note that Condition \ref{cond:1} encompasses special cases $p(x,y) > 0$ for all $(x,y) \in \mathcal{X}\times \mathcal{Y}$ and $p(x,y) = p(x) p(y)$ for all $(x,y) \in \mathcal{X} \times \mathcal{Y}$. The main implication is that whenever Condition \ref{cond:1} holds, each $x \in \mathcal{X}$ only belongs to a single hyperedge in the maximal hyperedge set $\Gamma_m^{0, f}$, stated in the following proposition.

\begin{prop} \label{prop:unique_clustering}
	If Condition \ref{cond:1} holds, hyperedges in $\Gamma_m^{0, f}$ do not overlap. In other words, if $x \in w_1, w_2$ for some $w_1, w_2 \in \Gamma_m^{0, f}$, then $w_1 = w_2$.
\end{prop}
\begin{IEEEproof}
	Without loss of generality, assume that $|w_1| \leq |w_2|$. If $w_1$ is singleton, then $x \in w_1, x \in w_2$ implies $w_1 \subset w_2$. This contradicts the fact that $w_1, w_2$ are in a maximal set. Hence, it is sufficient to assume that both $w_1, w_2$ are non-singleton.
	
	Suppose $w_1 \ne w_2$. Then, we can take $x_{w_1} \in w_1 \setminus w_2$, $x_{w_2} \in w_2 \setminus w_1$, and $x \in w_1 \cap w_2$. From Def.~\ref{def:e_characteristic_hypergraph}, one of the following cases must hold for each $y$:
	\begin{enumerate}
		\item[] \textsc{Case} 1: $f(x_{w_i},y) = f(x,y)$ with any probability $p(x,y), p(x_{w_i},y)$.
		\item[] \textsc{Case} 2: $f(x,y) \ne f(x_{w_i},y)$ with at least one of $p(x,y), p(x_{w_i}, y)$ being zero.
	\end{enumerate}
	Moreover, Condition \ref{cond:1} rules out the case when only one of $p(x,y), p(x_{w_i}, y)$ is zero in \textsc{Case} 2. Therefore, we only have
	\begin{enumerate}
		\item[] \textsc{Case} 1: $f(x_{w_i},y) = f(x,y)$ with any probability $p(x,y), p(x_{w_i},y)$.
		\item[] \textsc{Case} 2': $f(x,y) \ne f(x_{w_i},y)$ with $p(x,y), p(x_{w_i}, y)$ both being zero.
	\end{enumerate}	
	That is, $f(x,y), f(x_{w_1},y), f(x_{w_2},y)$ in \textsc{Case} 1 and 2' satisfy the SEC condition in Def.~\ref{def:e_characteristic_hypergraph}, which implies there is a hyperedge $w'$ such that $x_{w_1} \in w'$, $x_{w_2} \in w'$, and $x \in w'$. This contradicts the fact that $\Gamma_m^{0, f}$ is maximal. Hence, $w_1 = w_2$.
\end{IEEEproof}

Hence, under Condition \ref{cond:1}, given any $x \in \mathcal{X}$, quantizing $x$ into a unique hyperedge it belongs to and then using an optimal coding scheme in the side information setting, such as Slepian-Wolf coding with one rate being infinity, achieve the optimal rate.\footnote{Further, this scheme can be implemented in $\mathcal{O}(n\log{n})$ time overall since the quantization can be performed in $\mathcal{O}(n)$ time and the Slepian-Wolf coding is in $\mathcal{O}(n\log{n})$ time using polar codes \cite{Arikan2010} where $n$ is the blocklength.} The following two examples illustrate each case where the maximal hyperedges are overlapping or not.

\begin{figure}[t]
	\centering
	\subfloat[$p(x,y)$]{\begin{tabular}{c|ccc} \label{fig:ex1_a}
			\f{$X \backslash Y$} & \f{1} & \f{2} & \f{3} \\
			\hline
			\f{1} & \f{$\tfrac{1}{7}$} & \f{$\tfrac{1}{7}$} & \f{0} \\ 
			\f{2} & \f{$\tfrac{1}{7}$} & \f{$\tfrac{1}{7}$} & \f{$\tfrac{1}{7}$}\\
			\f{3} & \f{$\tfrac{1}{7}$} & \f{$\tfrac{1}{7}$} & \f{0} \\
	\end{tabular}}~~
	\subfloat[$f(x, y) = (x + y) \text{ mod } 2$]{~~~~\begin{tabular}{c|ccc} \label{fig:ex1_b}
			\f{$X \backslash Y$} & \f{1} & \f{2} & \f{3} \\
			\hline
			\f{1} & \f{0} & \f{1} & \f{?} \\
			\f{2} & \f{1} & \f{0} & \f{1} \\
			\f{3} & \f{0} & \f{1} & \f{?} \\
		\end{tabular}~~~~} \\
	\subfloat[Corresponding hypergraph $G^0$]{\includegraphics[width=1.2in]{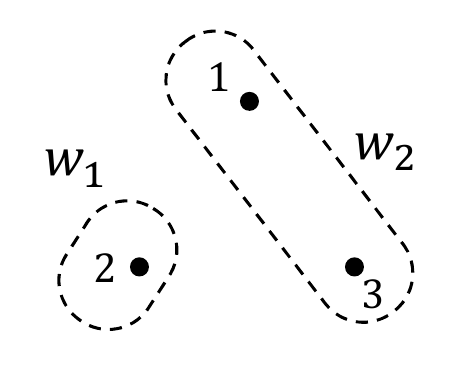} \label{fig:ex1_c}}
	\caption{Illustration of Example \ref{example:assumption}. In (b), ``?'' indicates that values could be any since such events do not occur.}
	\label{fig:ex_separate}
\end{figure}

\begin{ex} 	\label{example:assumption}
	Consider $(X,Y) \in \mathcal{X} \times \mathcal{Y}$, where $\mathcal{X} = \mathcal{Y} = \{1,2,3\}$, following $p(x,y)$ in Fig.~\ref{fig:ex1_a}. Let $f(x,y) = (x+y) \text{ mod } 2$ in Fig.~\ref{fig:ex1_b} be the function of interest. Then, the maximal $\epsilon$-characteristic hypergraph $G^{0,f}$ for $f$ is shown in Fig.~\ref{fig:ex1_c}, which has no overlapping hyperedges as it satisfies Condition \ref{cond:1}. Therefore, finding $\Gamma_m^{0, f}$, and then compressing $\Gamma_m^{0, f}$ via Slepian-Wolf encoding with respect to the induced $p(w,y)$ achieves the optimal rate.
\end{ex}

\begin{figure}[t]
	\centering
	\subfloat[$p(x,y)$]{\begin{tabular}{c|ccc} \label{fig:ex2_a}
			\f{$X \backslash Y$} & \f{1} & \f{2} & \f{3} \\
			\hline
			\f{1} & \f{0} & \f{$\tfrac{1}{6}$}  & \f{$\tfrac{1}{6}$} \\ 
			\f{2} & \f{$\tfrac{1}{6}$}  & \f{0}  & \f{$\tfrac{1}{6}$} \\
			\f{3} & \f{$\tfrac{1}{6}$}  & \f{$\tfrac{1}{6}$}  & \f{0} \\
	\end{tabular}}~~
	\subfloat[$f(x, y) = \mathbbm{1}_{\{ x > y \}}$]{~~~~\begin{tabular}{c|ccc} \label{fig:ex2_b}
			\f{$X\backslash Y$} & \f{1} & \f{2} & \f{3} \\
			\hline
			\f{1} & \f{?} & \f{0} & \f{0} \\
			\f{2} & \f{1} & \f{?} & \f{0}\\
			\f{3} & \f{1} & \f{1} & \f{?} \\
		\end{tabular}~~~~} \\
	\subfloat[Corresponding hypergraph $G^0$]{\includegraphics[width=1.2in]{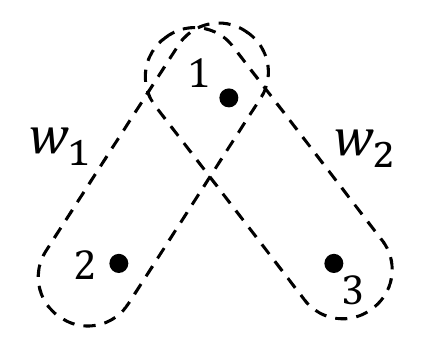} \label{fig:ex2_c}}
	\caption{Illustration of Example \ref{example:assumption_not_hold}. In (b), ``?'' indicates that values could be any since such events do not occur.}
	\label{fig:ex_non_separate}
\end{figure}

\begin{ex} \label{example:assumption_not_hold}
	Consider $(X,Y) \in \mathcal{X} \times \mathcal{Y}$, where $\mathcal{X} = \mathcal{Y} = \{1,2,3\}$ again, but with $p(x,y)$ in Fig.~\ref{fig:ex2_a} and $f(x, y) = \mathbbm{1}_{\{ x > y \}}$ in Fig.~\ref{fig:ex2_b}. As hyperedges overlap, $H(W|X)$ does not vanish. Hence, the optimal $p(w,x)$ needs to search over entire $p(w,x)$ space.
\end{ex}

\subsection{Partially Known Functions}
The purpose of approximate computing is to reduce the computation complexity for function $f$ that might be computationally demanding. To this end, in practice one may wish to know the performance when a simpler function is used; for instance, one may want to use a linear approximation $g$ to $f$. Another useful scenario for practical applications is that the encoder only has partial knowledge such that $f$ is $L$-Lipschitz, while the decoder knows $f$. We discuss the optimal rate bounds in such settings.

First, consider the case where the encoder and decoder both agree to use $g$ that is a $\delta$-approximation to $f$ as defined in Def.~\ref{def:ep_approx}. For instance, if there is a good (piecewise) linear or Taylor-based $\delta$-approximation to a computationally heavy function available to the encoder and decoder, then they might use $g$ rather than the actual $f$.
\begin{cor} \label{cor:Delta_1_ep}
	Let $g$ be a $\delta$-approximation to $f$. Then, for $\epsilon \ge \delta$, $R[\epsilon;f] \leq R[\epsilon-\delta; g]$. Further, the coding scheme achieving $R[\epsilon-\delta; g]$ also achieves the maximal distortion $\epsilon$ for $f$.
\end{cor}

\begin{IEEEproof}
	For $w \in \Gamma_m^{\epsilon-\delta, g}$, let $c_w$ be the center of the SEC of points $\{g(x,y): x \in w, p(x,y) > 0\}$. Then, by the definition of the maximal set of hyperedges,
	\begin{align*}
		\|g(x,y) - c_w\| \le \epsilon-\delta ~~~ \forall (x,y) \text{ such that } p(x,y) > 0.
	\end{align*}
	Further, $g$ is a $\delta$-approximation to $f$. That is, $\|f(x,y)-g(x,y)\| \le \delta$ for all $(x,y) \in \mathcal{X} \times \mathcal{Y}$. Therefore, as long as $x \in w \in \Gamma_m^{\epsilon-\delta, g}$, upon receiving the index of $W$, the decoder can declare $c_w$ such that
	\begin{align*}
		\|f(x,y) - c_w\| &\stackrel{(a)}{\le} \|f(x,y) - g(x,y)\| + \|g(x,y) - c_w\| \\
		&\le \delta + (\epsilon - \delta) = \epsilon,
	\end{align*}
	where (a) follows from the triangle inequality. Hence, the maximal $(\epsilon-\delta)$-characteristic hypergraph for $g$ ensures that the maximal distortion with respect to $f$ is less than or equal to $\epsilon$.
\end{IEEEproof}

Next, even the encoder does not need to know a specific function to compress for. Consider the class of $L$-Lipschitz functions.
\begin{cor} \label{cor:L_Lipschitz}
	Assume that $\mathcal{X} \times \mathcal{Y}$ form a metric space and the function of interest $f(x,y)$ is $L$-Lipschitz in $x$, i.e., $\| (x,y)-(x',y)\|\le c$ implies $\|f(x,y)-f(x',y)\| \le L c$ for all $y$, where $L, c$ are some positive constants. Also, let $\mathcal{F}_L=\{ f(x,y) \}$ be the class of $L$-Lipschitz functions in $x$ and $f_{\textsf{id}}(x,y) = x$ be the identity function in $x$. Then, the following holds.
	\begin{enumerate}
		\item[(i)] $R[\epsilon;f] \leq R[\epsilon/L; f_{\textsf{id}}]$
		
		\item[(ii)] Further assume that $\mathcal{X} \times \mathcal{Y}, \mathcal{Z}$ are vector spaces. Then, there exists a function $f^* \in \mathcal{F}_L$ such that the above inequality holds with equality. In addition, using the coding scheme based on the maximal $(\epsilon/L)$-characteristic hypergraph for $f_{\textsf{id}}$, an oracle that knows the true $f^*$ can reconstruct $f^*$ within tolerance level $\epsilon$.
	\end{enumerate}
\end{cor}
\begin{IEEEproof}
	The main idea is that if a set of $k$ points $\{x_{1}, \ldots, x_{k}\}$ has the SEC of radius $\epsilon/L$, then the set of points $\{f(x_{1},y), \ldots, f(x_{k},y)\}$ will have the SEC of radius less than or equal to $\epsilon$. Therefore, the maximal $\epsilon/L$-characteristic hypergraph for $f_{\textsf{id}}$ is also a (not necessarily maximal) $\epsilon$-characteristic hypergraph for $f$. When $\mathcal{X} \times \mathcal{Y}, \mathcal{Z}$ are vector spaces, a linear function having Lipschitz constant $L$ is an example of $f^*$.
\end{IEEEproof}

\section{Distributed coding for computing} \label{sec:distributed_coding}
In this section, we consider distributed coding for computing under maximal distortion. First, we give an existing inner bound based on the Berger-Tung inner bound and propose a hypergraph-based inner bound, then establish that our achievable scheme is as good as the Berger-Tung-based inner bound in the sum-rate sense. Later, we consider a special case and numerically compare our results with previous inner bounds in \cite{DoshiSME2010, FeiziM2014}, which shows that our achievable scheme outperforms them.

\subsection{Hypergraph-based Achievable Region} \label{subsec:achi_region}
We present the Berger-Tung inner bound under maximal distortion for coding for computing setting. The proof is immediate from that for standard distortion, thus omitted.

\begin{thm}[Berger-Tung inner bound \cite{Tung1978, Berger1977}] \label{thm:BT_inner} 
	For $f:\mathcal{X}_1 \times \mathcal{X}_2 \mapsto \mathcal{Z}$, let $\mathcal{R}_{\inn}[\epsilon] := \mathcal{R}_{\inn}(0;\dep) = \{(R_1, R_2) \in \mathbb{R}^2 \}$ be the set of rate pairs such that
	\begin{align*}
		R_1 &\ge I(X_1;U_1|U_2,Q), \\
		R_2 &\ge I(X_2;U_2|U_1,Q), \\
		R_1 + R_2 &\ge I(X_1, X_2;U_1, U_2|Q)
	\end{align*}
	for some joint pmf $p(q)p(u_1|x_1,q)p(u_2|x_2,q)$ with $|\mathcal{Q}| \leq 4$, $|\mathcal{U}_j| \leq |\mathcal{X}_j| + 4$, $j = 1,2$ and some function $z(u_1,u_2)$ such that $\mathbb{E} [ d_{\epsilon}(f(X_1,X_2), z(U_1,U_2)) ] = 0$, where $d_{\epsilon}(f(x_1, x_2), z) = \mathbbm{1}_{\{\|z - f(x_1,x_2)\| > \epsilon \}}$. Then, any rate pair in $\mathcal{R}_{\inn}[\epsilon]$ is achievable.
\end{thm}

The Berger-Tung-based inner bound is an optimization problem over $U_1, U_2$ and $z(u_1, u_2)$ satisfying the Markov chain and distortion constraint, which is non-constructive.

The next theorem implies that the hypergraph-based coding provides another achievable region obtainable with auxiliary random variables of finite cardinalities, i.e., constructive, and the achievable region belongs to $\mathcal{R}_{\inn}[\epsilon]$.

\begin{thm} \label{thm:hypergraph_achievable_region}
	For $f:\mathcal{X}_1 \times \mathcal{X}_2 \mapsto \mathcal{Z}$, let $(G_{m,1}^{\epsilon,f}, G_{m,2}^{\epsilon,f})$ be the pair of maximal $\epsilon$-characteristic hypergraphs in Def.~\ref{def:epsilon_hypergraph_pair}. Let $\mathcal{R}_{\G}[\epsilon]$ be the set of rate pairs $(R_1, R_2) \in \mathbb{R}_+^2$ such that
	\begin{align*}
		R_1 &\ge I(X_1;W_1|W_2,Q), \\
		R_2 &\ge I(X_2;W_2|W_1,Q), \\
		R_1 + R_2 &\ge I(X_1, X_2;W_1, W_2|Q)
	\end{align*}
	for some joint pmf $p(q)p(w_1|x_1,q)p(w_2|x_2,q)$ where $p(w_i|x_i,q)$ can be nonzero only when $x_i \in w_i \in \Gamma_{m,i}^{\epsilon, f}$. Then, $\mathcal{R}_{\G}[\epsilon] \subset \mathcal{R}_{\inn}[\epsilon]$.
\end{thm}
\begin{IEEEproof}
	Fix $p(x_1, x_2, w_1, w_2)$ satisfying $W_1-X_1-X_2-W_2$. For each pair of hyperedges $w_1 \in \Gamma_{m,1}^{\epsilon, f}, w_2 \in \Gamma_{m,2}^{\epsilon, f}$, take $z(w_1, w_2)$ as the center of the SEC of $\{f(x_1,x_2): x_1 \in w_1, x_2 \in w_2, p(x_1, x_2)>0\}$. Then, if $p(x_1, x_2)>0$, then $p(w_1, w_2) > 0$ for $w_1 \ni x_1, w_2 \ni x_2$ due to the definition of $\epsilon$-characteristic hypergraph. Therefore, $\| f(x_1,x_2) - z(w_1,w_2) \| \leq \epsilon$ for $x_1 \in w_1, x_2 \in w_2$ such that $p(x_1, x_2) > 0$, which implies $\mathbb{E}[\dep(X_1, X_2, z(W_1, W_2))] = 0$. 
	
	Observe that $W_1,W_2$, and $z$ from the above construction satisfy the constraints on the auxiliary random variables and reconstruction function in $\mathcal{R}_{\inn}[\epsilon]$. Therefore, repeating random codebook generation and error analysis of the Berger-Tung inner bound \cite{Tung1978}, \cite[Thm.~12.1]{ElGamalK2011}, the bounds on $R_1, R_2, R_1+R_2$ can be found. Since our hypergraph construction is a particular instance of $U_1, U_2$ in $\mathcal{R}_{\inn}[\epsilon]$, $\mathcal{R}_{\G}[\epsilon] \subset \mathcal{R}_{\inn}[\epsilon]$ holds.
\end{IEEEproof}

As noted, Thm.~\ref{thm:hypergraph_achievable_region} ensures that the hypergraph-based coding is computationally simpler than using general $U_1, U_2$. Thus, one may naturally wonder if it achieves the same rate performance as $\mathcal{R}_{\inn}[\epsilon]$. We give a positive answer to this at least for minimal achievable sum-rate, i.e., there is no rate degradation in minimal sum-rate. Let $R_{\G}^{\textsf{sum}}[\epsilon]:= \min_{(R_1, R_2) \in \mathcal{R}_{\G}[\epsilon]} R_1 + R_2$ and $R_{\inn}^{\textsf{sum}}[\epsilon]:= \min_{ (R_1, R_2) \in \mathcal{R}_{\inn}[\epsilon]} R_1 + R_2$. Then, the following holds.
\begin{thm} \label{thm:G<rect}
	The hypergraph-based Berger-Tung achievability does not incur rate loss in minimal achievable sum-rate. That is, $R_{\G}^{\textsf{sum}}[\epsilon] = R_{\inn}^{\textsf{sum}}[\epsilon]$.
\end{thm}
\begin{IEEEproof}
	The proof is in App.~\ref{app:pf_sum_rate}.
\end{IEEEproof}

\subsection{Special Case when Sources are Independent} \label{subsec:indep_source}

To further justify our hypergraph-based coding scheme, let us consider a simpler case where $X_1, X_2$ are independent. For this, the Berger-Tung inner and outer bounds \cite{Tung1978} meet; thus give the rate-distortion region. That is, the rate-distortion region is $\mathcal{R}_{\inn}[\epsilon]=\{(R_1,R_2)\}$ such that
\begin{align*}
	R_1 &\geq I(X_1;U_1|Q), \\
	R_2 &\geq I(X_2;U_2|Q)
\end{align*}
for some distributions $p(q)p(u_1|x_1,q)p(u_2|x_2,q)$ such that $$\mathbb{E}[d_{\epsilon}(f(X_1,X_2), \hat{f}(U_1,U_2))] = 0$$ for some function $\hat{f}$. However, it is still unknown how to achieve it in a constructive way as auxiliary random variables and recovered function values are general. To that end, consider the hypergraph rate region $\mathcal{R}_{\G}[\epsilon] = \{(R_1,R_2)\}$ such that
\begin{align*}
	R_1 &\geq I(X_1;W_1|Q), \\
	R_2 &\geq I(X_2;W_2|Q)
\end{align*}
for some joint pmf $p(q)p(w_1|x_1,q)p(w_2|x_2,q)$ where $p(w_i|x_i,q)$ can be nonzero only when $x_i \in w_i \in \Gamma_{m,i}^{\epsilon, f}$.
The following proposition justifies that our hypergraph-based coding scheme achieves the full rate-distortion region with smaller computational complexity.
\begin{prop} \label{prop:independent_source}
	If $X_1, X_2$ are independent, $\mathcal{R}_{\G}[\epsilon] = \mathcal{R}_{\inn}[\epsilon]$. That is, the hypergraph-based achievable region is tight.
\end{prop}
\begin{IEEEproof}
	As $\mathcal{R}_{\G}[\epsilon] \subset \mathcal{R}_{\inn}[\epsilon]$ in general, it is sufficient to show that $\mathcal{R}_{\G}[\epsilon] \supset \mathcal{R}_{\inn}[\epsilon]$. To this end, let $U_1,U_2$ be the auxiliary random variables that achieve a rate pair $(R_1, R_2) \in \mathcal{R}_{\inn}[\epsilon]$. Let $p(x_1, x_2, u_1, u_2)$ be the probability distribution over $(X_1, X_2, U_1, U_2)$ and define ${ \hat{w}_i(u_i) = \{x_i: p(x_i, u_i)>0\} }$ for $i \in \{1,2\}$. We can then define the Markov chain $(X_1,X_2) - (U_1,U_2) - (W_1,W_2)$ by
	\begin{align*}
		p(w_1, w_2|x_1, x_2, u_1, u_2) = \begin{cases}
			1 & \text{if } w_i = \hat{w}_i(u_i), i \in \{1,2\} \\
			0 & \text{otherwise. }
		\end{cases}
	\end{align*}
	
	The fact that $\mathbb{E} \left[ \mathbbm{1}_{\{ \|f(X_1,X_2) - Z(U_1,U_2)\| > \epsilon \}} \right] = 0$ implies $\|f(x_1,x_2) - z(u_1,u_2)\| \leq \epsilon$ for any $x_1, x_2, u_1, u_2$ with positive probability. In other words, for any $x_1,x_2$ such that $x_1 \in \hat{w}_1(u_1) = w_1, x_2\in \hat{w}_2(u_2) = w_2$, we have $\|f(x_1,x_2) - z(u_1,u_2)\| \leq \epsilon$. Hence, the radius of the SEC of points $\{f(x_1,x_2): x_1 \in w_1, x_2 \in w_2 \}$ is less than or equal to $\epsilon$. If $w_i$ is non-maximal, we can map $w_i$ to $\breve{w_i}$ that is maximal without increase in distortion. Thus, $X_i-U_i-W_i-\breve{W}_i$ holds and
	\begin{align*}
		I(X_i;U_i) \ge I(X_i;W_i) \ge I(X_i;\breve{W_i}).
	\end{align*}
	Induced probabilities $p(\breve{w}_i|x_i)$ through the Markov chain can span only a subset of all possible $p(\breve{w}_i|x_i)$, considering all $p(\breve{w}_i|x_i)$ over maximal hyperedges enlarges the achievable region. Finally, coded time-sharing over multiple pmfs via $q$ completes the proof.
\end{IEEEproof}

\begin{figure}[t]
	\centering
	\includegraphics[width=3.3in]{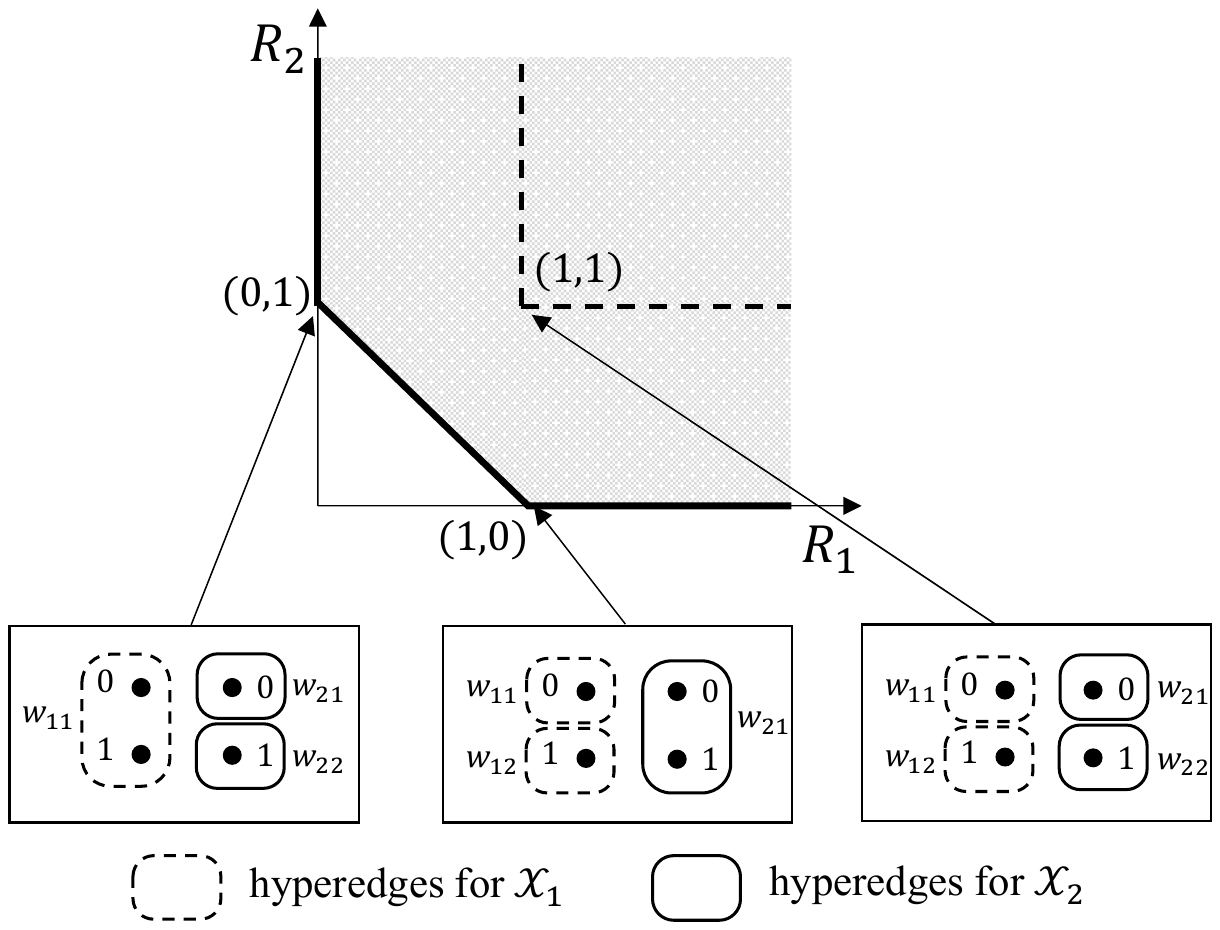}
	\caption{Rate-distortion region (shaded region) for Example \ref{ex:indep_source} at $\epsilon=0.5$ and corresponding hypergraphs. Sources are i.i.d.~$\text{Bern}(0.5)$ random variables and $f(x_1, x_2) = (x_1, x_2)$. The rate-distortion region is achievable by hypergraph-based coding in Thm.~\ref{thm:hypergraph_achievable_region}, while the achievable region by \cite{FeiziM2014} gives the dotted boundary that is suboptimal.}
	\label{fig:example_independent_source}
\end{figure}

For distributed lossy functional compression problems with independent sources, Feizi and M\'{e}dard \cite[Thm.~43]{FeiziM2014} also proposed an achievable rate-distortion region based on graph coloring, which is a generalization of Orlitsky and Roche's coding to lossy compression. At the core of their achievable scheme lies a hypergraph construction as well, called $D$-characteristic hypergraphs. We describe their coding scheme informally here.

First, \cite{FeiziM2014} defines hyperedges of $D$-characteristic hypergraph as follows. Let $i,j \in \{1,2\}$ with $i \ne j$.	
\begin{itemize}
	\item In \cite{FeiziM2014}: $w_i \subset \mathcal{E}_i$ is a set of $x_i$s such that the radius of SEC of $\{f(x_1, x_2): x_i \in w_i \}$ is less than or equal to $D$ for all $x_j \in \mathcal{X}_j$.
	
	\item Ours: $w_i \subset \mathcal{E}_i$ is a set of $x_i$s such that the radius of SEC of $\{f(x_1, x_2): x_i \in w_i, x_j \in w_j,  p(x_1,x_2) > 0 \}$ is less than or equal to $\epsilon$ for all $w_j \in \mathcal{E}_j$.
\end{itemize}
In other words, $D$-characteristic hypergraphs construct hyperedges of $x_i$ assuming $x_j$ can be anything, even though the decoder will have some information about $x_j$ via message $M_j$. Then, the coding scheme constructs a $D/2$-characteristic hypergraph for each input so that each input incurs a deviation up to $D/2$ each; the total deviation will be less than $D$. Then, the $D/2$-characteristic hypergraphs are encoded/decoded via Slepian-Wolf coding. Hence, each $D/2$-characteristic hypergraph \emph{separately} considers all possible values of the other input, while our approach \emph{jointly} considers the other simultaneously. Thus, our approach results in a more concise description. To be explicit, we compare the achievable rate-distortion regions for the following simple example.
\begin{ex} \label{ex:indep_source}
	Let $(X_1,X_2)$ be i.i.d.~$\text{Bern}(0.5)$ random variables and the function of interest be the identity function, i.e., $f(x_1, x_2) = (x_1, x_2) \in \mathbb{R}^2$. The distance metric is Euclidean. For $\epsilon = 0.5$, the achievable region by our hypergraph-based coding is shown in the shaded region in Fig.~\ref{fig:example_independent_source}, which turns out to be optimal by Prop.~\ref{prop:independent_source}. The points $(1,0)$ and $(0,1)$ are directly achievable by hypergraphs shown below, and the line between the two is by time-sharing. On the other hand, if \cite[Thm.~43]{FeiziM2014} is adopted, $1/4$-characteristic hypergraphs consist of singletons only; it is no different from directly compressing sources, and thus no rate reduction from functional compression.
\end{ex}

\section{Multiple Descriptions} \label{sec:further_demonstration}
\subsection{Multiple Description Coding} \label{sec:multiple_description_coding}
We consider multiple description coding for computing depicted in Fig.~\ref{fig:mdc}, where there are three distortion constraints. For this problem, the rate-distortion region is still unknown in general, and our focus is on the inner bound by El Gamal and Cover \cite{ElGamalC1982}. We show that the inner bound can be attained by our hypergraph coding.

The following is the El Gamal-Cover inner bound for the maximal distortion, which can be easily obtained from that for a general distortion.
\begin{thm}[El Gamal-Cover inner bound \cite{ElGamalC1982} for functional compression] \label{thm:mdc_elgamal_cover_inner}
	Let $\mathcal{R}_{\inn}[\epsilon_0, \epsilon_1, \epsilon_2] = \{ (R_1, R_2) \in \mathbb{R}^2 \} $ be the set of rate pairs such that
	\begin{align*}
		R_1 &\ge I(X; Z_1|Q), \\
		R_2 &\ge I(X; Z_2|Q), \\
		R_1 + R_2 &\ge I(X; Z_0,Z_1,Z_2|Q) + I(Z_1; Z_2|Q)
	\end{align*}
	for some conditional pmf {$p(q) p(z_0, z_1, z_2|x,q)$ with $|\mathcal{Q}| \leq 6$ such that $\mathbb{E} \left[ \|f_i(X) - Z_i\| > \epsilon_i \right] = 0$} for $i \in \{0,1,2\}$. Then, any rate pair in $\mathcal{R}_{\inn}[\epsilon_0, \epsilon_1, \epsilon_2]$ is achievable.
\end{thm}

The El Gamal-Cover inner bound provides the existence of achievable codes with the aid of exhaustive search over auxiliary random variable space, thus not constructive. Now in the following theorem, we provide a constructive hypegraph-based achievable region that reproduces the El Gamal-Cover inner bound. Let  $(G_{m,0}^{\epsilon_0, f_0}, G_{m,1}^{\epsilon_1, f_1}, G_{m,2}^{\epsilon_2, f_2} )$ be the triplet of the maximal $\epsilon_i$-characteristic hypergraphs, where each $G_{m,i}^{\epsilon_i, f_i} = (\mathcal{X}, \Gamma_{m,i}^{\epsilon_i, f_i})$ is constructed according to Defs.~\ref{def:e_characteristic_hypergraph} and \ref{def:max_characteristic_graph} with no side information, i.e., $Y_i = \emptyset$.

\begin{thm} \label{thm:mdc_hypergraph_achi_region}
	Let $\mathcal{R}_{\G}[\epsilon_0, \epsilon_1, \epsilon_2] = \{ (R_1, R_2) \in \mathbb{R}^2 \}$ be the set of rate pairs such that
	\begin{align*}
		R_1 &\ge I(X;W_1|Q), \\
		R_2 &\ge I(X;W_2|Q), \\
		R_1 + R_2 &\ge I(X;W_0,W_1,W_2|Q) + I(W_1; W_2|Q)
	\end{align*}
	for some conditional pmf $p(q)p(w_0,w_1,w_2|x,q)$ with $|\mathcal{Q}|\le 6$ where $p(w_i|x,q)$ can be nonzero only when $x \in w_i \in \Gamma_{m,i}^{\epsilon_i, f_i}$ for $i \in \{0,1,2\}$. Then, $\mathcal{R}_{\G}[\epsilon_0, \epsilon_1, \epsilon_2] = \mathcal{R}_{\inn}[\epsilon_0, \epsilon_1, \epsilon_2]$.
\end{thm}
\begin{IEEEproof}
	Let us first prove ``$\subset$'' direction. Consider $x \in w \in \Gamma_{m,i}^{\epsilon_i, f_i}$ for $i \in \{0,1,2\}$. Letting $z_{w}$ be the center of the SEC of points $\{f_i(x): x \in w\}$, by the construction of the maximal $\epsilon_i$-characteristic hypergraph, we know that $\| f_i(x)-z_{w} \|\le \epsilon_i$ for all $x \in w$. Hence, all $W$s in $\Gamma_{m, i}^{\epsilon_i, f_i}$ satisfy the constraints on the auxiliary random variables in $\mathcal{R}_{\inn}[\epsilon_0, \epsilon_1, \epsilon_2]$. As the rate-bound expressions are the same, the hyperedges and the center of them are a particular instance of $Z_0, Z_1, Z_2$ and $\hat{f}_i(Z_i)$. Therefore, $\mathcal{R}_{\G}[\epsilon_0, \epsilon_1, \epsilon_2]$ is achievable by El Gamal-Cover's achievability and $\mathcal{R}_{\G}[\epsilon_0, \epsilon_1, \epsilon_2] \subset \mathcal{R}_{\inn}[\epsilon_0, \epsilon_1, \epsilon_2]$.
	
	To show ``$\supset$'' direction, consider some $(Z_0, Z_1, Z_2)$ with $p(z_0, z_1, z_2|x)$ that satisfy the constraints of $\mathcal{R}_{\inn}[\epsilon_0, \epsilon_1, \epsilon_2]$. Letting $\hat{w}_i(z_i) = \{x: p(z_i,x) > 0\} ~ \text{for $i \in \{0,1,2\}$}$, we can define the Markov chain $(W_0, W_1, W_2) - (Z_0, Z_1, Z_2) - X$ by taking
	\begin{align} \label{eqn:mc2}
		&p(w_0,w_1,w_2|z_0,z_1,z_2,x) = p(w_0,w_1,w_2|z_0,z_1,z_2) \nonumber \\
		&= \begin{cases}
			1, & \text{if $w_0 = \hat{w}_0(z_0),w_1 = \hat{w}_1(z_1),w_2 = \hat{w}_2(z_1)$} \\
			0, & \text{otherwise.}
		\end{cases}
	\end{align}
	Then, as $W_i$ is determined by $Z_i$, the Markov chain $W_i-Z_i-X$ holds as well. 
	
	To show $p(w_i|x,q)$ can be nonzero only when $X \in W_i$, note that
	\begin{align*}
		p(x, w_i) &= \sum_{z_i} p(x, z_i, w_i) = \sum_{z_i} p(x, z_i) p(w_i|z_i, x) \\
		&= \sum_{z_i} p(x, z_i) \mathbbm{1}_{\{w_i = \hat{w}_i(z_i)\}}.
	\end{align*}
	Hence, $p(x, w_i)>0$ implies that there is a $z_i$ such that $w_i = \hat{w}_i(z_i)$ and $p(x, z_i)>0$, which in turn implies $p(w_i|x)$ is nonzero only when $x \in \hat{w}_i(z_i) = w_i$ from the definition of $\hat{w}_i$.
	
	To show $w_i \in \Gamma_{m,i}^{\epsilon_i, f_i}$, take an arbitrary $x \in w_i$. By its construction, there is a $z_i$ such that $p(x, z_i) = p(x)p(z_i|x) > 0$ and moreover, $\|f_i(x)-z_i\|\le \epsilon_i$. Since $x$ is arbitrary, $\|f_i(x)-z_i\|\le \epsilon_i$ holds for all $x \in w_i$. That is, the radius of the SEC is less than or equal to $\epsilon_i$. In other words, $\{w_i\}$ is a hyperedge set of $\epsilon_i$-characteristic hypergraph, possibly non-maximal.
	
	Now, consider the rate bounds. The bounds $I(X;W_i) \le I(X;Z_i)$ follow from the Markov chains $W_i-Z_i-X$. The Markov chain $(W_0, W_1, W_2) - (Z_0, Z_1, Z_2) - X$ implies $I(X;W_0, W_1, W_2) \le I(X;Z_0, Z_1, Z_2)$. Also, $Z_1$ and $Z_2$ are independent conditioned on $X$, we have $W_1-Z_1-X-Z_2-W_2$. It implies $I(W_1;W_2) \le I(Z_1;Z_2)$.
	
	If $\{w_i\}$ is non-maximal, then we can further map $w_i$ to $\breve{w}_i$ so that $\breve{w}_i$ is maximal. In other words, $X-W_i-\breve{W}_i$ holds, which gives $I(X;W_i) \ge I(X;\breve{W}_i)$. The mapping also yields two Markov chains $X - (W_0, W_1, W_2) - (\breve{W}_0, \breve{W}_1, \breve{W}_2)$ and $\breve{W}_1 - W_1 - X - W_2 - \breve{W}_2$. It means
	\begin{align*}
		&I(X;W_1, W_2, W_3) + I(W_1;W_2) \\
		&\ge I(X;\breve{W}_1, \breve{W}_2, \breve{W}_3) + I(\breve{W}_1;\breve{W}_2).
	\end{align*}	
	Since $\breve{W}_i$s obtained from the mapping yields $\Gamma_{m,i}^{\epsilon_i, f_i}$, but does not span full $p(\breve{w}_i|x)$ over the maximal hyperedge set. Therefore, optimizing over all possible $p(\breve{w}_i|x)$ further minimizes the rate bounds.	Generalizing to coded time-sharing completes the proof of ``$\supset$'' direction.	
\end{IEEEproof}

A special case of multiple description coding is the setting where decoder $2$ does not exist. It is called the successive refinement problem, where message $M_2$ plays a role of additional information on $X^n$ that helps decoder $0$ refine the recovery of $f_0(X)$. For this setting, it is known that the El Gamal-Cover inner bound in Thm.~\ref{thm:mdc_elgamal_cover_inner} is tight \cite{EquitzC1991, Rimoldi1994}. Since hypergraph-based achievability reproduces the El Gamal-Cover inner bound, it also gives the rate-distortion region for functional successive refinement under maximal distortion.
\begin{cor}
	Let $\mathcal{R}_{\G}[\epsilon_0, \epsilon_1] = \{ (R_1, R_2) \in \mathbb{R}^2 \}$ be the set of rate pairs such that
	\begin{align*}
		R_1 &\ge I(X;W_1), \\
		R_1 + R_2 &\ge I(X;W_0,W_1)
	\end{align*}
	for some conditional pmf $p(w_0,w_1|x)$ where $p(w_i|x)$ can be nonzero only when $x \in w_i \in \Gamma_{m,i}^{\epsilon_i, f_i}$ for $i \in \{0,1\}$. Then, $\mathcal{R}_{\G}[\epsilon_0, \epsilon_1]$ is the rate-distortion region for the successive refinement problem.
\end{cor}

\subsection{Cascade Multiple Description Coding} \label{sec:cascade_mdc}
In this section, we discuss cascade multiple description coding with three nodes, depicted in Fig.~\ref{fig:cascade_mdc}, for which the (non-constructive) rate-distortion region is from Yamamoto \cite{Yamamoto1981}. We demonstrate that our hypergraph-based coding scheme achieves the same rate-distortion region as Yamamoto's with smaller search complexity, i.e., optimal with smaller complexity.

Thm.~\ref{thm:yamamoto_region} gives the rate-distortion region under maximal distortion, which is a special case of general distortion \cite{Yamamoto1981}.
\begin{thm}[\!\!\cite{Yamamoto1981}] \label{thm:yamamoto_region}
	Let $\mathcal{R}[\epsilon_1, \epsilon_2] = \{(R_1, R_2) \in \mathbb{R}_+^2\}$ such that
	\begin{align*}
		R_1 &\ge I(X_1, X_2;Z_1, Z_2), \\
		R_2 &\ge I(X_1, X_2;Z_2)
	\end{align*}
	for some conditional pmf $p(z_1,z_2|x_1, x_2)$ satisfying $\mathbb{E} \left[ \mathbbm{1}_{\{ \|f_i(X_i) - Z_i\| > \epsilon_i \}} \right] = 0$ for $i \in \{1,2\}$. Then, $\mathcal{R}[\epsilon_1, \epsilon_2]$ is the rate-distortion region, i.e., any rate pair in $\mathcal{R}[\epsilon_1, \epsilon_2]$ is achievable, while no rate pair outside $\mathcal{R}[\epsilon_1, \epsilon_2]$ is achievable.
\end{thm}

Similar to the previous settings, $\mathcal{R}[\epsilon_1, \epsilon_2]$ still relies on exhaustive searching over the auxiliary variable space to span a rate pair on the boundary. In Thm.~\ref{def:cmdc_hyper_achi_region}, we provide a hypegraph-based achievable region $\mathcal{R}_{\G}[\epsilon_1, \epsilon_2]$ under maximal distortion, which is constructive and tight. For this, consider a pair of the maximal $\epsilon_i$-characteristic hypergraphs, $(G_{m,1}^{\epsilon_1, f_1}, G_{m,2}^{\epsilon_2, f_2} )$, where each $G_{m,i}^{\epsilon_i, f_i} = (\mathcal{X}_i, \Gamma_{m,i}^{\epsilon_i, f_i})$ is constructed according to Defs.~\ref{def:e_characteristic_hypergraph} and \ref{def:max_characteristic_graph} with no side information, i.e., $Y_i = \emptyset$.	

\begin{thm} \label{def:cmdc_hyper_achi_region}
	Let $\mathcal{R}_{G}[\epsilon_1, \epsilon_2] = \{(R_1, R_2) \in \mathbb{R}_+^2\}$ such that
	\begin{align*}
		R_1 &\ge I(X_1, X_2;W_1,W_2), \\
		R_2 &\ge I(X_1, X_2;W_2)
	\end{align*}
	for some conditional pmf $p(w_1,w_2|x_1, x_2)$ where $p(w_i|x_i)$ can be nonzero only when $x \in w_i \in \Gamma_{m,i}^{\epsilon_i, f_i}$ for $i \in \{1,2\}$. Then, $\mathcal{R}_{G}[\epsilon_1, \epsilon_2] = \mathcal{R}[\epsilon_1, \epsilon_2]$.
\end{thm}
\begin{IEEEproof}
	Let us first prove ``$\subset$'' direction. Fix an arbitrary $p(w_1,w_2|x_1, x_2)$ over $(\Gamma_{m,1}^{\epsilon_1, f_1}, \Gamma_{m,2}^{\epsilon_2, f_2})$. Then, by the definition of a pair of maximal $\epsilon_i$-characteristic hypergraphs, we know that if $x_i \in w_i \in \Gamma_{m,i}^{\epsilon_i, f_i}$, then $\|f_i(x_i) - c(w_i)\| \leq \epsilon_i$ with $c(w_i)$ being the center of the SEC of points $\{f_i(x_i): x_i \in w_i\}$. Hence, $W_i$ in $\mathcal{R}_{\G}[\epsilon_1, \epsilon_2]$ satisfies the constraints on the auxiliary random variables in $\mathcal{R}[\epsilon_1, \epsilon_2]$, and then we have found a particular instance of such auxiliary random variables. Yamamoto's achievability gives the same rate bounds on $R_1$ and $R_2$ implying that $\mathcal{R}_{\G}[\epsilon_1, \epsilon_2] \subset \mathcal{R}[\epsilon_1, \epsilon_2]$.
	
	To show ``$\supset$'' direction, fix an arbitrary $Z_1, Z_2$ with $p(z_1, z_2|x_1, x_2)$ satisfying $\mathbb{E} \left[ \mathbbm{1}_{\{ \|f_i(X_i) - Z_i\| > \epsilon_i \}} \right] = 0$. Then, let $	\hat{w}_i(z_i) = \{x_i: p(z_i,x_i) > 0\} ~ \text{ for } i \in \{1,2\}$ and take the Markov chain $(W_1, W_2) - (Z_1, Z_2) - (X_1, X_2)$ as
	\begin{align} \label{eqn:cmc2}
		&p(w_1,w_2|z_1,z_2,x_1, x_2) = p(w_1,w_2|z_1,z_2) \nonumber \\
		&= \begin{cases}
			1, & \text{if $w_1 = \hat{w}_1(z_1),w_2 = \hat{w}_2(z_2)$}\\
			0, & \text{otherwise.}
		\end{cases}
	\end{align}
	From construction \eqref{eqn:cmc2}, we also have that $p(w_i|z_i,x_1, x_2) = \mathbbm{1}_{\{w_i = \hat{w}_i(z_i)\}}$ for $i \in \{1,2\}$, hence Markov chains $(X_1, X_2) - Z_i - W_i, i=1,2$ hold.
	
	To show $p(w_i|x_i)$ can be nonzero only when $x_i \in w_i$, note that 
	\begin{align*}
		p(x_i, w_i) = \sum_{z_i} p(x_i, z_i, w_i) = \sum_{z_i} p(x_i, z_i) \mathbbm{1}_{\{w_i = \hat{w}_i(z_i)\}}.
	\end{align*}	
	Hence, $p(x_i, w_i) > 0$ implies that there is at least one $z_i$ such that $w_i = \hat{w}_i(z_i)$ and $p(x_i, z_i)>0$, which in turn implies $x_i \in \hat{w}_i(z_i) = w_i$ from $\hat{w}$.
	
	To show $w_i \in \Gamma_{m,i}^{\epsilon_i, f_i}$ for $i \in \{1,2\}$, take an arbitrary $x_i \in w_i$. As $Z_i$ satisfies the distortion constraint, we know $\|f_i(x_i)-z_i\|\le \epsilon_i$ for $z_i$ such that $p(x_i, z_i) > 0$. Since $x_i \in w_i$ is arbitrary, $\|f_i(x_i)-z_i\|\le \epsilon_i$ holds for all $x_i \in w_i$. That is, the radius of the SEC of $\{ f_i(x_i):x_i \in w_i \}$ is less than or equal to $\epsilon_i$. Hence $\{w_i\}$ forms a set of hyperedges of an $\epsilon_i$-characteristic hypergraph.
	
	Using the data processing inequality, the inequality $I(X_1, X_2;Z_1,Z_2) \ge I(X_1, X_2; W_1, W_2)$ follows from Markov chain $(X_1, X_2) - (Z_1, Z_2) - (W_1, W_2)$. Further, $I(X_1, X_2;Z_2) \ge I(X_1, X_2;W_2)$ can be obtained from the Markov chain $(X_1, X_2) - Z_2 - W_2$. If $w_i$ is non-maximal, we can map it to $\breve{w}_i \in \Gamma_{m, i}^{\epsilon_i, f_i}$ implying Markov chains $(X_1, X_2)-(W_1, W_2)-(\breve{W}_1, \breve{W}_2)$ and $(X_1, X_2) - W_i - \breve{W}_i$. This leads to $I(X_1, X_2;W_1,W_2) \ge I(X_1, X_2;\breve{W}_1,\breve{W}_2)$ and $I(X_1, X_2;W_2) \ge I(X_1, X_2;\breve{W}_2)$. Since $p(\breve{w_1}, \breve{w_2}|x_1, x_2)$ induced from the Markov chains in general cannot span full test channels over the maximal hyperedges, further minimizing over $\Gamma_{m,i}^{\epsilon_i, f_i}$ completes the proof.
\end{IEEEproof}

\section{Markov Sources with Sparse Hypergraphs} \label{sec:sparse_markov}
This section illustrates the computational benefit of hypergraph-based coding when the source is Markov and the corresponding hypergraph is \emph{sparse}, which will be formally defined later. For simplicity, the focus is on the point-to-point case. Before proceeding, let us review the literature on compression for a general stationary first-order Markov source over a finite alphabet $\mathcal{X}$ and discuss its computational complexity via na\"{i}ve numerical optimization algorithms.

\subsection{Background}
To illustrate the computational complexity arising from the Markovity, we assume the target function is the identity and the distortion of interest is general. Then, the rate-distortion function is known in multi-letter form:
\begin{align}
	R(D) = \lim_{n \to \infty} \frac{1}{n} \min_{\mathbf{p} \in \mathbf{P}} I(X^n; \hat{X}^n), \label{eq:multi_letter}
\end{align}
where $\mathbf{p}$ is a conditional probability in the set $\mathbf{P} = \{ p(\hat{x}^n | x^n ): \frac{1}{n}\sum_{t=1}^n \mathbb{E}[d(X_t, \hat{X}_t)] \le D \}$.

The single-letter expression is unknown, and thus much effort has been devoted to obtaining computable bounds on \eqref{eq:multi_letter}. The best upper bound, i.e., achievable rate, is by Jalali and Weissman \cite{JalaliW2007},\footnote{The paper mainly focuses on the symmetric binary case, but it can be immediately extended to a general Markov source.}
\begin{align}
	R(D) \le \frac{k}{k+1} \widetilde{R}_k(\widetilde{D}) + \frac{H(X_{k+2} | X_1)}{k+1}, \label{eq:k_letter}
\end{align}
where $\widetilde{R}_k, \widetilde{D}$ will be specified soon. To see how the bound can be attained, fix $k \in \mathbb{N}$ and let $S_i := X_{(i-1)(k+1)+1}$ and supersymbol $U_i$ be middle symbols between $S_i$ and $S_{i+1}$:
\begin{align*}
	\underbrace{X_1}_{=: S_1}, \underbrace{X_2, \ldots, X_{k+1}}_{=:U_1}, \underbrace{X_{k+2}}_{=: S_2}, \ldots.
\end{align*}
Then, the encoding scheme is two-stage:
\begin{enumerate}
	\item Compress $\{S_1, S_2, \ldots\}$ losslessly. As the source is stationary, it needs $\frac{H(S_{i+1}|S_i)}{k+1} = \frac{H(X_{k+2}|X_1)}{k+1}$ rate. This step leads to the second term in \eqref{eq:k_letter}.
	
	\item Compress $\{U_1, U_2, \ldots\}$ within distortion $\widetilde{D}=\frac{k+1}{k} D$. As $S_{t}, S_{t+1}$ give some information about $U_t$, $\{U_t\}$ is a mixture process of $|\mathcal{X}|^2$ types of i.i.d.~processes. This step leads to the first term in \eqref{eq:k_letter}.
\end{enumerate}
Hence, the achievability converts the infinite-letter problem \eqref{eq:multi_letter} to a $k$-letter one with
\begin{align*}
	\widetilde{R}_k(\widetilde{D}) = \sum_{i,j \in \mathcal{X}^2} \Pr[X_1=i, X_{k+2}=j] \widetilde{R}_k^{i,j}(\widetilde{D}),
\end{align*}
where $\widetilde{R}_k^{i,j}(\widetilde{D})$ is the rate-distortion function for i.i.d.~supersymbol $U$ drawn from $\Pr[x_2^{k+1}|(X_1, X_{k+2})=(i,j)]$. It is asymptotically tight with $k$ \cite{JalaliW2007}.

The main difficulty is now in Step 2), which is in $k$-letter form. Consequently, to obtain $\widetilde{R}_k$, a numerical optimization algorithm must be run over test channels $p(\hat{x}_2^{k+1}|x_2^{k+1})$ satisfying the distortion constraint. However, the test channel is over $p(\hat{x}_2^{k+1}|x_2^{k+1}) \in \mathbb{R}^{|\mathcal{X}|^k} \times \mathbb{R}^{|\mathcal{X}|^k}$, which is $|\mathcal{X}|^{2k}$-dimensional. That means, if $|\mathcal{X}|, k$ are large, a na\"{i}ve optimization algorithm (e.g., Blahut-Arimoto) that takes all $|\mathcal{X}|^{2k}$ transition pairs into account cannot be run, or it needs additional efforts to reduce the complexity. For a lossy functional compression problem such that we desire to approximate $f(x_t)$, the dimension of the space of $\hat{u}=\hat{x}_2^{k+1}$ could be infinite since $\hat{x}_t$ is allowed to take a continuous value to approximate. Our hypergraph-based scheme under maximal distortion \emph{preprocesses} the symbols and reduces the complexity in an efficient and interpretable manner.

\subsection{Markov Sources with Sparse Hypergraphs}
Let $(Z_1, Z_2, \ldots)$ be a (not necessarily Markov) process over a finite alphabet $\mathcal{Z}$. Then, we call the process \emph{$s$-sparse} if the transition probability is sparse in the sense that
\begin{align*}
	|\{z_{t+1}: p(z_{t+1}|z_t) > 0\}| \le s ~ \text{for all } t, z_t.
\end{align*}

Recall the achievable scheme by Jalali and Weissman \cite{JalaliW2007} and extend it to the maximal distortion with target function $f$. Using hypergraphs, we quantize (preprocess) each symbol $x_t$ to a hyperedge $w_t$ and find the best test channel for a block of length $k$ in Step 2), i.e., $p(w_2^{k+1}|x_2^{k+1})$. Then, the rate-distortion function for a Markov source can be upper-bounded as follows.
\begin{align}
	R[\epsilon] &\le \frac{k}{k+1} \widetilde{R}_k[\epsilon] + \frac{H(X_{k+2} | X_1)}{k+1} \nonumber \\
	&= \frac{k}{k+1} \widetilde{R}_k(0; \dep) + \frac{H(X_{k+2} | X_1)}{k+1}, \label{eq:Markov_ub_maximal}
\end{align}
where $\widetilde{R}_k[\epsilon] = \sum_{i,j \in \mathcal{X}^2} \Pr[X_1=i, X_{k+2}=j] \widetilde{R}_k^{i,j}[\epsilon]$. To be precise, this upper bound can be attained as follows.
\begin{enumerate}
	\item Compress $\{S_1, S_2, \ldots\}$ losslessly. It needs $\frac{H(X_{k+2}|X_1)}{k+1}$ rate. The decoder maps recovered $\hat{S}_i$ to a hyperedge that $\hat{S}_i$ can belong to in a predetermined manner.
	
	\item Note that $(\hat{S}_{i}, \hat{S}_{i+1}) = (S_{i}, S_{i+1})$ reveals the type of $X_{(i-1)(k+1)+2}^{i(k+1)}$. Per each type, find the best mapping from $X_{(i-1)(k+1)+2}^{i(k+1)}$ to $W_{(i-1)(k+1)+2}^{i(k+1)}$, where $W_t$ is a hyperedge of the maximal $\epsilon$-characteristic hypergraph. It needs additional $\frac{k}{k+1} \widetilde{R}_k[\epsilon]$ bits.
\end{enumerate}
The next corollary is about the complexity of solving $\widetilde{R}_k[\epsilon]$ numerically when the hypergraph process is $s$-sparse.

\begin{cor} \label{cor:sparse_dimension}
	Consider a stationary Markov source $(X_1, X_2, \ldots)$ and its maximal $\epsilon$-characteristic hypergraph $G_{m}^{\epsilon,f}$. If a process of hyperedges $(W_1, W_2, \ldots)$ induced from the Markov source and $f$ is $s$-sparse, i.e., $|\{w_{t+1}: p(w_{t+1}|w_t) > 0\}| \le s=s(\epsilon) ~ \text{for all } t$, then an admissible test channel for \eqref{eq:Markov_ub_maximal} is equal to or less than $(s|\mathcal{X}|)^k$-dimensional.
\end{cor}
\begin{IEEEproof}
	To simplify indices, let $X_2^{k+1}, W_2^{k+1}$ be generic copies of $X_{(i-1)(k+1)+2}^{i(k+1)},W_{i(k+1)+2}^{(i-1)(k+1)}$, respectively. Also, let $G_m^{\epsilon, f} = (\mathcal{X}, \Gamma_m^{\epsilon, f})$ be the maximal hypergraph.
	
	Fix $x_2^{k+1}$. Although there are total $|\Gamma_m^{\epsilon, f}|^k$ number of $W_2^{k+1}$ including ones with zero probability, there are only at most $s$ probable values of $w_3$ given $w_2$ according to the sparsity assumption. Similarly, given $(w_2, w_3)$, there are only at most $s$ probable values of $w_4$. Repeating this, we know that there are at most $s^k$ probable values of $W_2^{k+1}$. It means the space of $p(w_2^{k+1}|x_2^{k+1})$ is indeed $s^k$-dimensional if $x_2^{k+1}$ is fixed. Since there are at most $|\mathcal{X}|^{k}$ probable values of $x_2^{k+1}$, $p(w_2^{k+1}|x_2^{k+1})$ is in the $(s^k \cdot |\mathcal{X}|^k)$-dimensional space at most.
\end{IEEEproof}

The following simple example illustrates the dimension reduction when $X$ is a birth-death process.
\begin{figure}[t]
	\centering
	\includegraphics[width=3.0in]{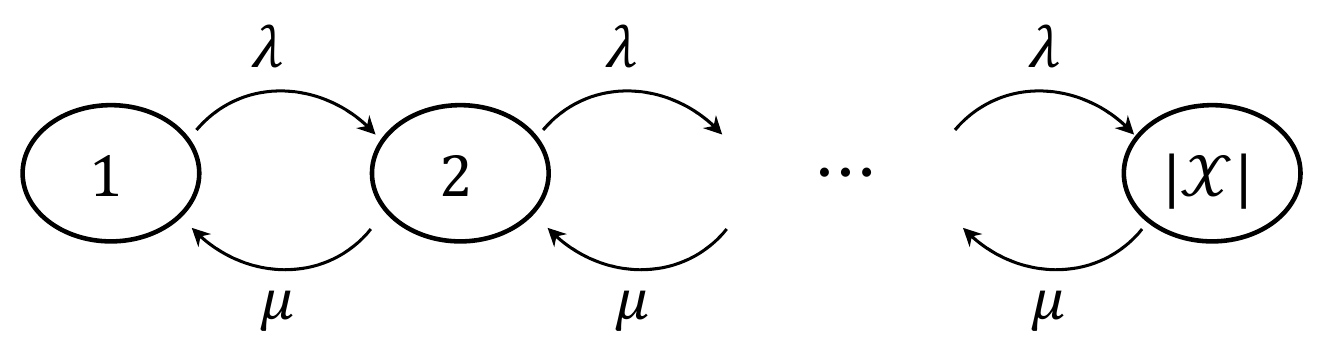}
	\caption{The birth-death process in Example \ref{ex:birth_death}.}
	\label{fig:birth_death_process}
\end{figure}
\begin{ex} \label{ex:birth_death}
	Consider a birth-death process, depicted in Fig.~\ref{fig:birth_death_process}, consisting of state space $\mathcal{X} = \{1,2,\ldots, |\mathcal{X}|\}$ with birth and death probabilities $\lambda, \mu >0$ respectively. The target function is identity, i.e., $f(x_t) = x_t$, and the desired tolerance level of the maximal distortion is $\epsilon = 0.5$ under the Euclidean distance. Then, a na\"{i}ve optimization algorithm searches for the optimal $p(\hat{x}_2^{k+1} | x_2^{k+1}) \in \mathbb{R}^{|\mathcal{X}|^{2k}}$ to obtain $\widetilde{R}_2[0.5]$.
	
	Note that under tolerance $\epsilon=0.5$, two consecutive states belong to the same hyperedge because $\{f(i), f(i+1)\} = \{i, i+1\}$ can be enclosed by a circle of radius $0.5$ centered at $i+\tfrac{1}{2}$. Therefore, the set of maximal hypergraphs for a single symbol $x_t$ is $\Gamma_m^{\epsilon, f} = \{w_i, i=1,2,\ldots, |\mathcal{X}|-1\}$ with $w_i = \{i, i+1\}$, and it is sufficient for the decoder to receive $w_i$ to recover $f(x_t)$ within tolerance $\epsilon$.
	
	We then know that by the birth-death law the hyperedge process is $3$-sparse, which implies the hypergraph-based coding reduces the dimension of a test channel's space to $(3|\mathcal{X}|)^k$. Note that it is much smaller than $|\mathcal{X}|^{2k}$ if $|\mathcal{X}| \gg 3$.
\end{ex}

\section{Conclusion} \label{sec:conclusion}
Inspired by modern applications and approximate computing, this work considers functional compression problems under maximal distortion. Leveraging a geometric property of the maximal distortion, we propose a hypergraph-based coding scheme that explicitly specifies optimal or good auxiliary random variables. The rate-distortion performance is demonstrated in several settings for which the resulting performance is optimal or at least reproduces (a part of) known inner bounds. Moreover, the complexity of optimizing test channels can be largely reduced by our hypergraph-based coding, which is illustrated for Markov sources such that an induced hypergraph is sparse.

While we have demonstrated that our hypergraph-based coding is useful for functional compression for approximate computing, much work remains. For example, in Sec.~\ref{sec:distributed_coding}, we only considered sum-rate equivalence for distributed coding since the Markov property used for the proof of Thm.~\ref{thm:G<rect} is inapplicable to proving individual rate bounds. Hence, we may need to develop a different technical tool. Another direction one may wonder is how to capture \emph{common} information explicitly using hypergraphs. For the multiple description coding problem, the Zhang-Berger inner bound \cite{ZhangB1987} extends the El Gamal-Berger inner bound by adopting the common information. Likewise, we expect that capturing common information further improves the achievable region in Thm.~\ref{thm:mdc_hypergraph_achi_region}.

\section*{Acknowledgement}
We thank S. Roy, H. Yadav, A. Deshmukh, A. Magesh, and I. Jain for valuable discussions.

\appendices
\section{Portion of Proof of Theorem \ref{thm:coding_for_computing_side_info}} \label{app:pf_cfc_side_info}
We need to check whether this $W$ satisfies $W-X-Y$. Since Markov chain $W-V-(X, Y)$ holds, we have $p(w, x, y|v) = p(w|v)p(x, y | v)$. Marginalizing out $y$ term at both sides, we also have a new Markov chain $W-V-X$. Then,
\begin{align*}
	p(w|x, y) &= \sum_v p(v|x,y) p(w|v,x,y) \stackrel{(a)}{=} \sum_v p(v|x) p(w|v) \\
	&\stackrel{(b)}{=} \sum_v p(v|x) p(w|v, x) = \sum_v p(v,w|x) = p(w|x),
\end{align*}
where (a) follows from two Markov chains $V-X-Y$ and $W-V-(X, Y)$ and (b) follows from $W-V-X$. The equality implies Markov chain $W-X-Y$ holds.

The remaining step of the proof is to show $I(X;W|Y) \leq I(X;V|Y)$. To show this, we can obtain $W-V-Y$ from $W-V-(X, Y)$ by marginalization. Also, $W-(V,Y)-X$ holds since
\begin{align*}
	p(w, x, y|v) &\stackrel{(c)}{=} p(w|v) p(x, y|v) \\
	\implies p(y|v) p(w, x|v, y) &= p(w|v) p(y|v) p(x|v, y) \\
	\implies p(w, x|v, y) &= p(w|v) p(x|v, y) \\
	& \stackrel{(d)}{=} p(w|v,y) p(x|v,y).
\end{align*}
where (c) follows from $W-V-(X,Y)$ and (d) follows from $W-V-X$.
Then,
\begin{align*}
	I(X;W|Y) &= H(X|Y) - H(X|W, Y) \\
	&\stackrel{(e)}{\le} H(X|Y) - H(X|W, V, Y) \\
	&\stackrel{(f)}{=} H(X|Y) - H(X|V, Y) = I(X;V|Y),
\end{align*}
where (e) holds since conditioning reduces entropy and (f) follows from $W-(V,Y)-X$. This completes the proof.

\section{Proof of Theorem \ref{thm:G<rect}} \label{app:pf_sum_rate}
As $\mathcal{R}_{\G}[\epsilon] \subset \mathcal{R}_{\inn}[\epsilon]$ from Thm.~\ref{thm:hypergraph_achievable_region}, ``$\ge$'' direction is immediate.

To show ``$\le$'' direction, fix $(U_1,U_2)$ and $z(u_1,u_2)$ that achieve $R_{\inn}^{\textsf{sum}}[\epsilon]$. We will show that there exists a pair of $\epsilon$-characteristic hypergraphs induced by $(U_1,U_2), z(u_1,u_2)$ and then consider the identity for the pair of maximal $\epsilon$-characteristic hypergraph.

Letting $p(x_1,x_2,u_1,u_2)$ be the pmf over $(X_1,X_2,U_1,U_2)$, define $\hat{w}_1, \hat{w}_2$ as $\hat{w}_i(u_i) = \{x_i: p(x_i,u_i)>0\}$. Also, take a probability
\begin{align} \label{eqn:mc1}
	p(w_1,w_2|x_1,x_2,u_1,u_2) = 
	\begin{cases}
		1, & \text{if } w_i = \hat{w}_i(u_i), i \in \{1,2\} \\
		0, & \text{otherwise.}
	\end{cases}
\end{align}
Then, a Markov chain $(X_1,X_2) - (U_1,U_2) - (W_1,W_2)$ holds, which in turn implies $I(X_1,X_2;W_1,W_2) \leq I(X_1,X_2;U_1,U_2)$ holds by the data processing inequality. We will prove three claims: (i) $X_1 \in W_1 \in G_{1}^{\epsilon, f}$ and $X_2 \in W_2 \in G_{2}^{\epsilon, f}$, i.e., such $W_1, W_2$ form (not necessarily maximal) hyperedges of $\epsilon$-characteristic hypergraphs, (ii) the Markov chain $W_1-X_1-X_2-W_2$ holds, and (iii) ``$\le$'' direction holds for the pair of maximal hypergraphs.

(i) If $p(x_i,\hat{w}_i)>0$, then from \eqref{eqn:mc1}, there exists $u_i$ such that $w_i = \hat{w}_i(u_i)$ and $p(u_i,x_i) > 0$. Thus, $x_i \in  w_i = \hat{w}_i(u_i)$.  Fix any $w_1, w_2$ such that $p(w_1)>0, p(w_2)>0$. Then, there are $u_1, u_2$ such that $w_1 = \hat{w}_1(u_1)$, $w_2 = \hat{w}_2(u_2)$. Further, if $x_1 \in w_1$, $x_2 \in w_2$, then $p(x_1,u_1)>0$ and $p(x_2,u_2)>0$ for such $u_1, u_2$. In this case, $p(x_1,x_2) > 0$ implies $p(u_1,u_2,x_1,x_2)>0$ by Markov chain $U_1-X_1-X_2-U_2$. If $p(x_1,x_2) > 0$, $p(x_1,u_1)>0$, and $p(x_2,u_2)>0$, then using $U_1-X_1-X_2-U_2$, we have $p(u_1,u_2,x_1,x_2)>0$. Since we already know that $\mathbb{E} \left[ \mathbbm{1}_{\{ \|f(X_1,X_2) - z(U_1,U_2)\| > \epsilon \}} \right] = 0$, it follows that $\|f(x_1,x_2) - z(u_1,u_2)\| \leq \epsilon$ for any $x_1, x_2, u_1, u_2$ with positive probability. In other words, for any $(x_1,x_2)$ with $p(x_1,x_2)>0, x_1\in \hat{w}_1(u_1) = w_1, x_2\in \hat{w}_2(u_2) = w_2$, we have $\|f(x_1,x_2) - \hat{f}(u_1,u_2)\| \leq \epsilon$. Hence, the radius of the SEC of the set of points $\{f(x_1,x_2): x_1 \in w_1, x_2 \in w_2, p(x_1,x_2)>0\}$ is less than or equal to $\epsilon$. Thus, $w_1, w_2$ are hyperedges of $\epsilon$-characteristic hypergraphs, not necessarily maximal.

(ii) From two Markov chains $(X_1,X_2) - (U_1,U_2) - (W_1,W_2)$ and $U_1-X_1-X_2-U_2$, we have $W_1 - X_1 - X_2 - W_2$ since 
\begin{align*}
	&p(x_1, x_2, w_1, w_2) = p(x_1, x_2) p(w_1, w_2|x_1,x_2) \\
	&= p(x_1,x_2) \left( \sum_{u_1}\mathbbm{1}_{\{w_1=\hat{w}_1(u_1)\}} p(u_1|x_1) \right) \\
	&~~~~~~~~~~~~~~~~~ \times \left( \sum_{u_2}\mathbbm{1}_{\{w_2=\hat{w}_2(u_2)\}} p(u_2|x_2) \right) \\
	&= p(x_1,x_2)p(w_1|x_1)p(w_2|x_2).
\end{align*}

(iii) To show ``$\le$'' direction for the pair of maximal $\epsilon$-characteristic hypergraphs, consider a (possibly stochastic) map from a non-maximal hyperedge $W$ to maximal one $\breve{W}$ that contains the non-maximal hyperedge. Then, it forms a Markov chain $(X_1, X_2) - (U_1, U_2) - (W_1, W_2) - (\breve{W}_1, \breve{W}_2)$, which in turn implies that $I(X_1, X_2 ; W_1, W_2) \ge I(X_1, X_2 ; \breve{W}_1, \breve{W}_2)$ by the data processing inequality. Since $\breve{W}_1, \breve{W}_2$ taken through the Markov chain can only spans a subset of all possible $p(\breve{w}_i|x_i)$ over the maximal hyperedges, further taking $p(\breve{w}_i|x_i)$ directly over $\Gamma_{m,i}^{\epsilon, f}$ only reduces the rate, which also means a Markov chain $\breve{W}_1 - X_1 - X_2 - \breve{W}_2$ where $\breve{W}_i \in \Gamma_{m,i}^{\epsilon, f}$. In other words, $I(X_1, X_2 ; \breve{W}_1, \breve{W}_2) \ge R_{\G}^{\textsf{sum}}[\epsilon]= \min_{(R_1, R_2) \in \mathcal{R}_{\G}[\epsilon]} R_1 + R_2$. Combining inequalities concludes $R_{\G}^{\textsf{sum}}[\epsilon] \le R_{\inn}^{\textsf{sum}}[\epsilon]$.

\newpage

\begin{IEEEbiographynophoto}{Sourya~Basu} received the Bachelor of Technology degree in electrical engineering from the Indian Institute of Technology Kanpur, Kanpur, India, in 2017. He is currently pursuing the Ph.D.~degree in electrical and computer engineering with the University of Illinois at Urbana-Champaign. His research interests lie in the area of geometric deep learning, information theory, and natural language processing.
\end{IEEEbiographynophoto}

\begin{IEEEbiographynophoto} {Daewon~Seo} received the B.S.~(summa cum laude) and M.S.~degrees in electrical engineering from the Korea Advanced Institute of Science and Technology (KAIST), Daejeon, South Korea, in 2008 and 2010, respectively. After the M.S.~degree, he was with KAIST Institute and LG Electronics, South Korea, from 2010 to 2011 and from 2011 to 2014, respectively. He received the Ph.D.~degree in electrical and computer engineering from the University of Illinois Urbana-Champaign in 2019. He was with the Ming Hsieh department of electrical and computer engineering, the University of Southern California, Los Angeles, CA, USA, in 2019 and the department of electrical and computer engineering, the University of Wisconsin-Madison, Madison, WI, USA from 2019 to 2020. He is currently an assistant professor in the department of electrical engineering and computer science at Daegu Gyeongbuk Institute of Science and Technology (DGIST), Daegu, South Korea. His research interests include machine learning systems, information theory, and statistical inference.
\end{IEEEbiographynophoto}

\begin{IEEEbiographynophoto}{Lav R. Varshney} (S'00--M'10--SM'15) received the B.S.~degree (magna cum laude) in electrical and computer engineering with honors from Cornell University, Ithaca, New York, in 2004. He received the S.M., E.E., and Ph.D.~degrees, all in electrical engineering and computer science, from the Massachusetts Institute of Technology, Cambridge, in 2006, 2008, and 2010, where his theses received the E.~A.~Guillemin Thesis Award and the J.-A.~Kong Award Honorable Mention.
	
He is an associate professor in the Department of Electrical and Computer Engineering and the Coordinated Science Laboratory, University of Illinois at Urbana-Champaign, with further appointments in computer science, industrial engineering, neuroscience, digital agriculture, and personalized nutrition. He is also affiliated with the Discovery Partners Institute of the University of Illinois System.  He is currently a White House Fellow, serving in the National Security Council, Cyber and Emerging Technology, Washington, DC, USA.  During 2019--2020, he was a principal research scientist at Salesforce Research, Palo Alto, CA, USA.   During 2010--2013, he was a research staff member at the IBM Thomas J.~Watson Research Center, Yorktown Heights, NY, USA. His research interests include information and coding theory.
\end{IEEEbiographynophoto}

\vfill
\end{document}